\documentclass[preprint,12pt,review]{elsarticle}

\usepackage{amssymb}
\usepackage{amsthm}
\usepackage{amsmath}
\usepackage{mathrsfs}
\usepackage{graphicx}
\usepackage{epstopdf}
\usepackage{float}
\usepackage{caption}
\usepackage{subcaption}
\usepackage{bm}
\usepackage{amssymb} 
\usepackage{extarrows} 
\usepackage{bbm}
\usepackage{mathrsfs}
\usepackage{cleveref}
\usepackage{soul}
\usepackage{siunitx}
\usepackage{tabu} 
\usepackage{longtable}
\usepackage{xcolor} 
\usepackage{comment} 
\usepackage[margin=2.3cm]{geometry}
\biboptions{sort&compress} 

\renewcommand{\sin}[1]{\, \mbox{sin}\left( #1 \right) }
\renewcommand{\cos}[1]{\, \mbox{cos}\left( #1 \right) }
\renewcommand{\exp}[1]{\, \mbox{exp}\left( #1 \right) }

\journal{International Journal of Fatigue}

\makeatletter
\def\@author#1{\g@addto@macro\elsauthors{\normalsize%
    \def\baselinestretch{1}%
    \upshape\authorsep#1\unskip\textsuperscript{%
      \ifx\@fnmark\@empty\else\unskip\sep\@fnmark\let\sep=,\fi
      \ifx\@corref\@empty\else\unskip\sep\@corref\let\sep=,\fi
      }%
    \def\authorsep{\unskip,\space}%
    \global\let\@fnmark\@empty
    \global\let\@corref\@empty  
    \global\let\sep\@empty}%
    \@eadauthor={#1}
}
\makeatother

\begin{document}

\begin{frontmatter}



\title{A phase field model for hydrogen-assisted fatigue}


\author{Alireza Golahmar\fnref{DTU}}

\author{Philip K. Kristensen\fnref{DTU}}

\author{Christian F. Niordson\fnref{DTU}}

\author{Emilio Mart\'{\i}nez-Pa\~neda\corref{cor1}\fnref{IC}}
\ead{e.martinez-paneda@imperial.ac.uk}

\address[DTU]{Department of Mechanical Engineering, Technical University of Denmark, DK-2800 Kgs. Lyngby, Denmark}

\address[IC]{Department of Civil and Environmental Engineering, Imperial College London, London SW7 2AZ, UK}

\cortext[cor1]{Corresponding author.}

\begin{abstract}
We present a new theoretical and numerical phase field-based formulation for predicting hydrogen-assisted fatigue. The coupled deformation-diffusion-damage model presented enables predicting fatigue crack nucleation and growth for arbitrary loading patterns and specimen geometries. The role of hydrogen in increasing fatigue crack growth rates and decreasing the number of cycles to failure is investigated. Our numerical experiments enable mapping the three loading frequency regimes and naturally recover Paris law behaviour for various hydrogen concentrations. In addition, \emph{Virtual} S-N curves are obtained for both notched and smooth samples, exhibiting a good agreement with experiments.  

\end{abstract}


\end{frontmatter}


\section{Introduction}
\label{Sec:Intro} 

There is a growing interest in understanding and optimising the fatigue behaviour of metals in the presence of hydrogen (see, e.g., \cite{Martin2013,Colombo2015,Yamabe2017,Peral2019,Shinko2019,Ogawa2020} and Refs. therein). Two aspects have mainly motivated these endeavours. Firstly, hydrogen-assisted cracking is a well-known concern in the transport, construction, defence and energy sectors. Hydrogen is ubiquitous and significantly reduces the ductility, strength, toughness and fatigue crack growth resistance of metallic materials, with the problem being exacerbated by the higher susceptibility of modern, high-strength alloys \cite{Gangloff2003}. Secondly, hydrogen is seen as the energy carrier of the future, fostering a notable interest in the design and prognosis of infrastructure for hydrogen transportation and storage \cite{Murakami2006,Gangloff2012}. In the majority of these applications, susceptible components are exposed to alternating mechanical loads and thus being able to predict the synergistic effects of hydrogen and fatigue damage is of utmost importance.\\

Significant progress has been achieved in the development of computational models for hydrogen-assisted fracture. Dislocation-based methods \cite{AM2016,Burns2016a}, weakest-link approaches \cite{Novak2010,Ayas2014}, cohesive zone models \cite{Serebrinsky2004,Yu2017,Elmukashfi2020}, gradient damage theories \cite{Anand2019} and phase field fracture formulations \cite{CMAME2018,Duda2018,Wu2020b,JMPS2020} have been presented to predict the nucleation and subsequent growth of hydrogen-assisted cracks. Multi-physics phase field fracture models have been particularly successful, demonstrating their ability to capture complex cracking conditions, such as nucleation from multiple sites or the coalescence of numerous defects, in arbitrary geometries and dimensions \cite{TAFM2020c,CS2020}. However, the surge in modelling efforts experienced in the context of monotonic, static fracture has not been observed in fatigue. Hydrogen can influence the cyclic constitutive behaviour \cite{Castelluccio2018,Hosseini2021}, reduce the number of cycles required to initiate cracks \cite{Esaklul1983,Esaklul1983b} and, most notably, accelerate fatigue crack growth \cite{AM2020,Shinko2021}. Predicting the significant reduction in fatigue life observed in the presence of hydrogen requires capturing how hydrogen elevates crack growth rates, which is dependent on the hydrogen content, the material susceptibility to embrittlement, the diffusivity of hydrogen and the loading amplitude and frequency, among other factors. Given the complexity and higher computational demands of fatigue damage, it is not surprising that the role of hydrogen in augmenting fatigue crack growth rates has been predominantly assessed from an experimental viewpoint, with a few exceptions \cite{Moriconi2014,EFM2017}. Moreover, the success of phase field formulations in predicting hydrogen-assisted static fracture has not been extended to fatigue yet.\\

In this work, we present the first phase field model for hydrogen-assisted fatigue. The main elements of the coupled deformation-diffusion-fatigue formulation presented are: (i) a thermodynamically-consistent extension of Fick's law of mass diffusion, (ii) a fatigue history variable and associated degradation function, (iii) a phase field description of crack-solid interface evolution, (iv) a penalty-based formulation to update environmental boundary conditions, and (v) an atomistically-inspired relation between the hydrogen content and the fracture surface energy. This novel variational framework is numerically implemented in the context of the finite element method and used to model hydrogen-assisted fatigue in several boundary value problems of particular interest. Firstly, the paradigmatic benchmark of a cracked square plate is modelled to quantify the dependency of the number of cycles to failure on the hydrogen content. Secondly, a boundary layer approach is used to gain insight into the competing role of loading frequency and hydrogen diffusivity. We show how the model captures the main experimental trends; namely, the sensitivity of fatigue crack growth rates to the loading frequency and the environment. The Paris law, and its sensitivity to hydrogen, are naturally recovered. Finally, \emph{Virtual} S-N curves are computed for both smooth and notched samples, exhibiting a promising agreement with experimental data. The remainder of the paper is organized as follows. Section \ref{Sec:Theory} presents the theoretical framework and provides details of the finite element implementation. In Section \ref{Sec:Results}, the performance of the proposed modelling framework is benchmarked against several representative numerical examples as well as relevant experimental measurements. Finally, concluding remarks are given in Chapter \ref{Sec:Conclusions}.

\section{A phase field theory for hydrogen-assisted fatigue}
\label{Sec:Theory}

We present a theoretical and numerical framework for modelling hydrogen assisted fatigue. Our formulation is grounded on the phase field fracture method, which has gained notable traction in recent years. Applications include battery materials \cite{Zhao2016,Mesgarnejad2019}, composites \cite{Quintanas-Corominas2020a,CST2021}, ceramics \cite{Carollo2018,Li2021}, shape memory alloys \cite{CMAME2021}, functionally graded materials \cite{CPB2019,Kumar2021} and both ductile \cite{McAuliffe2015,Borden2016} and embrittled \cite{IJP2021} metals. The success of phase field fracture methods is arguably twofold. First, phase field provides a robust computational framework to simulate complex cracking phenomena in arbitrary geometries and dimensions. Secondly, it provides a variational platform for Griffith's energy balance \cite{Francfort1998,Bourdin2000}. Thus, consider a cracked elastic solid with strain energy density $\psi (\bm{\varepsilon})$. Under prescribed displacements, the variation of the total potential energy of the solid $\mathcal{E}$ due to an incremental increase in crack area d$A$ is given by
\begin{equation}
\frac{\text{d} \mathcal{E}}{\text{d} A} = \frac{\text{d} \psi (\bm{\varepsilon})}{\text{d} A} + \frac{\text{d} W_c}{\text{d} A}  = 0,
\end{equation}

\noindent where $W_c$ is the work required to create new surfaces and $\bm{\varepsilon}$ is the strain tensor. The fracture resistance of the solid is given by the term $\text{d}W_c/\text{d}A$, also referred to as the material toughness or critical energy release rate $G_c$. A pre-existing crack will grow when the energy stored in the material is high enough to overcome $G_c$. Griffith's minimality principle can be formulated in a variational form as follows
\begin{equation}\label{Eq:Pi}
\mathcal{E} = \int_\Omega \psi \left( \bm{\varepsilon} \right) \text{d} V + \int_\Gamma   G_c \, \text{d} \Gamma \, .
\end{equation}

\noindent Arbitrary cracking phenomena can be predicted based on the thermodynamics of fracture, provided one can computationally track the crack surface $\Gamma$. The phase field paradigm is key to tackling the challenge of predicting the evolution of the crack surface topology. The crack-solid interface is described by means of an auxiliary variable, the phase field $\phi$, which takes distinct values in each of the phases and varies smoothly in between. This implicit representation of an evolving interface has proven to be useful in modelling other complex interfacial phenomena, such as microstructural evolution \cite{Provatas2011} or corrosion \cite{JMPS2021}. In the context of fracture mechanics, the phase field $\phi$ resembles a damage variable, taking values of 0 in intact material points and of 1 inside the crack. Thus, upon a convenient constitutive choice for the crack surface density function $\gamma$, the Griffith functional (\ref{Eq:Pi}) can be approximated by means of the following regularised functional:
\begin{equation}\label{Eq:Piphi}
\mathcal{E}_\ell = \int_\Omega \big[ g \left( \phi \right) \psi_0 \left( \bm{\varepsilon} \right) + G_c \gamma \left( \phi, \ell \right) \big] \,  \text{d} V  = \int_\Omega \bigg[  \left( 1 - \phi \right)^2 \psi_0 \left( \bm{\varepsilon} \right) + G_c  \left( \frac{\phi^2}{2 \ell} + \frac{\ell}{2} | \nabla \phi |^2 \right) \bigg] \,  \text{d} V  \, .
\end{equation}

\noindent Here, $\ell$ is a length scale parameter that governs the size of the fracture process zone, $\psi_0$ denotes the strain energy density of the undamaged solid and $g(\phi)$ is a degradation function. It can be shown through Gamma-convergence that $\mathcal{E}_\ell$ converges to $\mathcal{E}$ when $\ell \to 0^+$ \cite{Chambolle2004}.

Now, let us extend this framework to incorporate fatigue damage and hydrogen embrittlement. Define a \emph{degraded} fracture energy $\mathcal{G}_d$ that is a function of the hydrogen concentration $C$ and a fatigue history variable $\bar{\alpha}$, such that
\begin{equation}
    \mathcal{G}_d = f_C \left( C \right) f_{\bar{\alpha}} \left( \bar{\alpha} \right) G_c \,
\end{equation}

\noindent where $f_C$ and $f_{\bar{\alpha}}$ are two suitably defined degradation functions to respectively incorporate hydrogen and fatigue damage, as described later. Replacing $G_c$ by $\mathcal{G}_d$, taking the variation of the functional (\ref{Eq:Piphi}) with respect to $\delta \phi$, and applying Gauss' divergence theorem renders the following phase field equilibrium equation,
\begin{equation}\label{eq:StrongFormEq}
    \mathcal{G}_d \left( \frac{\phi}{\ell} - \ell \nabla^2 \phi \right) - 2 \left( 1 - \phi \right) \psi_0 =0
\end{equation}

Considering the homogeneous solution to (\ref{eq:StrongFormEq}) provides further insight into the role of the phase field length scale $\ell$. Thus, in a 1D setting, consider a sample with Young's modulus $E$, subjected to a uniaxial stress $\sigma=g \left( \phi \right) E \varepsilon$; the homogeneous solution for the stress reaches a maximum at the following critical strength:
\begin{equation}\label{eq:Sc}
    \sigma_c = \left( \frac{27 E \mathcal{G}_d}{256 \ell} \right)^{1/2} \, .
\end{equation}

\noindent Hence, $\ell$ can be seen not only as a regularising parameter but also as a material property that defines the material strength. This enables phase field models to predict crack nucleation and naturally recover the transition flaw size effect \cite{Tanne2018,PTRSA2021}. 

\subsection{Hydrogen degradation function}

We proceed to provide constitutive definitions for the degradation functions. The dramatic drop in fracture resistance observed in metals exposed to hydrogen is captured by taking inspiration from atomistic insight. As discussed elsewhere \cite{Serebrinsky2004,CMAME2018}, DFT calculations of surface energy degradation with hydrogen coverage $\theta$ exhibit a linear trend, with the slope being sensitive to the material system under consideration. Thus, a quantum mechanically informed degradation law can be defined as follows,
\begin{equation}\label{eq:f_H}
    f_C = 1 - \chi \theta \,\,\,\,\,\,\,\,\,\,\,\,\,\, \text{with} \,\,\,\,\,\,\,\,\,\,\,\,\,\, \theta=\frac{C}{C+\exp{-\Delta g_b^0/(\mathcal{R}T)}}
\end{equation}

\noindent where $\chi$ is the hydrogen damage coefficient, which is taken in this study to be $\chi=0.89$, as this provides the best fit to the DFT calculations by Jiang and Carter in iron \cite{Jiang2004a,CMAME2018}. Also, the second part of (\ref{eq:f_H}) makes use of the Langmuir–McLean isotherm to estimate, as dictated by thermodynamic equilibrium, the hydrogen coverage $\theta$ at decohering interfaces as a function of the bulk concentration $C$, the universal gas constant $\mathcal{R}$, the temperature $T$, and the associated binding energy $\Delta g_b^0$. Here, we follow Serebrinsky \textit{et al.} \cite{Serebrinsky2004} and assume $\Delta g_b^0=30$ kJ/mol, as is commonly done for grain boundaries. These specific choices are based on the assumption of a hydrogen assisted fracture process governed by interface decohesion. However, we emphasise that the phase field framework for hydrogen assisted fatigue presented is general and can accommodate any mechanistic or phenomenological interpretation upon suitable choices of $f_C$.

\subsection{Fatigue degradation function}

Fatigue damage is captured by means of a degradation function $f_{\bar{\alpha}} \left( \bar{\alpha} \right)$, a cumulative history variable $\bar{\alpha}$ and a fatigue threshold parameter $\alpha_T$. Following the work by Carrara \textit{et al.} \cite{Carrara2020}, two forms of $f_{\bar{\alpha}} \left( \bar{\alpha} \right)$ are considered:
\begin{align} \label{eq:FdegAsy}
    f_{\bar{\alpha}}(\bar{\alpha})&=\left\{\begin{array}{ll}{1} & {\text { if } \quad \bar{\alpha} \leq \alpha_{T}} \\ {\left(\dfrac{2 \alpha_{T}}{\bar{\alpha}+\alpha_{T}}\right)^{2}} & {\text { if } \quad \bar{\alpha} > \alpha_{T}}\end{array}\right. \quad\quad (\text{Asymptotic})\\[3mm]
    f_{\bar{\alpha}}(\bar{\alpha})&=\left\{\begin{array}{ll}
    1 & \text { if } \quad \bar{\alpha} \leq \alpha_{T} \\
    {\left[1-\kappa \log \left(\dfrac{\bar{\alpha}}{\alpha_{T}}\right)\right]^{2}} & \text { if } \quad \alpha_{T} \leq \bar{\alpha} \leq \alpha_{T} 10^{1 / \kappa} \quad\quad \text { (Logarithmic) } \\
    0 & \text { if } \quad \bar{\alpha} \geq \alpha_{T} 10^{1 / \kappa}
\end{array}\right. \label{eq:FdegLog}
\end{align}

\noindent where $\kappa$ is a material parameter that governs the slope of the logarithmic function. For simplicity, the asymptotic function will be generally used in our numerical experiments unless otherwise stated. The fatigue history variable $\bar{\alpha}$ evolves in time $t$ as follows,
\begin{equation}
    \bar{\alpha}(t)=\int_{0}^{t} \mathrm{H}(\alpha \dot{\alpha})|\dot{\alpha}| \, \mathrm{d}t \, ,
\end{equation}

\noindent where $\mathrm{H}(\alpha \dot{\alpha})$ is the Heaviside function, such that $\bar{\alpha}$ only grows during loading. Finally, consistent with our energy balance, the cumulative fatigue variable is defined as $\bar{\alpha}=g \left( \phi \right) \psi_0$. 

\subsection{Coupled deformation-diffusion-fracture problem}

The hydrogen and fatigue damage framework presented is coupled to the solution of the displacement field, as given by the balance of linear momentum:
\begin{equation}\label{eq:BalanceLinearMomentum}
    \nabla \cdot \bm{\sigma} + \mathbf{b} = \bm{0} \, ,
\end{equation}

\noindent and mass transport,
\begin{equation}\label{eq:massTransport}
    \dot{C} + \nabla \cdot \mathbf{J} = 0 \, .
\end{equation}

\noindent Here, $\bm{\sigma}$ is the Cauchy stress tensor, $\bm{b}$ is the body force vector, and $\mathbf{J}$ is the hydrogen flux. In relation to the mechanical problem, linear elastic material behaviour is assumed, with the strain energy density given as $\psi_0=\frac{1}{2} \, \bm{\varepsilon} : \bm{\mathcal{C}} : \bm{\varepsilon} $, where $\bm{\mathcal{C}}$ is the fourth order elasticity tensor. The hydrogen transport problem is characterised by the following definition of the chemical potential,
\begin{equation}\label{eq:chemicalpotential}
    \mu = \mu_0 + \mathcal{R}T \, \text{ln} \frac{\theta}{1-\theta} - \bar{V}_H \sigma_H \,
\end{equation}
 
\noindent where $\mu_0$ denotes the chemical potential in the standard state and $\bar{V}_{H}$ is the partial molar volume of hydrogen in solid solution. Our numerical examples are focused on iron-based materials and consequently $\bar{V}_{H} = 2000$ mm$^3$/mol. It must be emphasised that the hydrostatic stress $\sigma_H$ lowers the chemical potential, increasing the hydrogen solubility as a result of lattice dilatation and thus attracting hydrogen to areas of high volumetric strains, such as crack tips. Finally, the hydrogen flux is related to $\nabla \mu$ through the following linear Onsager relation,
\begin{equation}
    \mathbf{J} = - \frac{DC}{\mathcal{R}T} \nabla \mu \, ,
\end{equation}

\noindent where $D$ is the hydrogen diffusion coefficient. The role of microstructural trapping sites in slowing diffusion can be accounted for by considering $D$ to be the effective diffusion coefficient (as opposed to the lattice one). Also, as shown in Ref. \cite{IJP2021} in the context of static fracture, the framework can readily be extended to capture the influence of dislocation traps, which evolve with mechanical load.

\subsection{Numerical implementation}

The weak forms of Eqs. (\ref{eq:StrongFormEq}), (\ref{eq:BalanceLinearMomentum}) and (\ref{eq:massTransport}) are discretised and solved using the finite element method. In addition, the following features enrich our numerical implementation. Firstly, damage irreversibility is enforced by means of a history field that satisfies the Kuhn-Tucker conditions \cite{Miehe2010a}. Secondly, damage under compressive fields is prevented by adopting a tension-compression split of the strain energy density, together with a hybrid implementation \cite{Ambati2015}. Two approaches are considered, the volumetric-deviatoric split by Amor \textit{et al.} \cite{Amor2009} and the spectral decomposition by Miehe \textit{et al.} \cite{Miehe2010a}; the former is generally used unless otherwise stated. Thirdly, the system of equations is solved with a staggered approach that converges to the monolithic result upon controlling the residual norm \cite{Seles2019a}. Finally, a penalty approach is adopted to implement \emph{moving} chemical boundary conditions, by which the diffusion-environment interface evolves as dictated by the phase field crack \cite{CS2020,Renard2020,JMPS2020}. 

\section{Results} 
\label{Sec:Results}

The predictive capabilities of the model are demonstrated through the following numerical experiments. Firstly, in Section \ref{sec:4_Plate}, we validate our numerical implementation in the absence of hydrogen and extend it to demonstrate how the model can capture the role of hydrogen in accelerating crack growth rates. Secondly, in Section \ref{sec:4_BoundaryLayer}, we use a boundary layer formulation to gain insight into hydrogen-assisted fatigue crack growth under small scale yielding conditions. Stationary and propagating cracks are modelled to shed light on the sensitivity of the crack tip hydrogen concentration to the fatigue frequency and compute Paris law coefficients for various hydrogen contents. Also, crack growth rates versus loading frequency regimes are mapped. Thirdly, we examine the fracture and fatigue behaviour of notched components in Section \ref{sec:4_NotchedBar}, computing \emph{Virtual} S-N curves for various hydrogenous environments. Finally, in Section \ref{sec:4_ExperimentalComparison} we compare model predictions with fatigue experiments on smooth samples, observing a very good agreement. Two materials are considered, with samples being exposed either to air or to high pressure hydrogen gas. 

\subsection{Cracked square plate subjected to fatigue in a hydrogenous environment}
\label{sec:4_Plate}

\begin{figure}[H]
\centering
\includegraphics[width=1\linewidth]{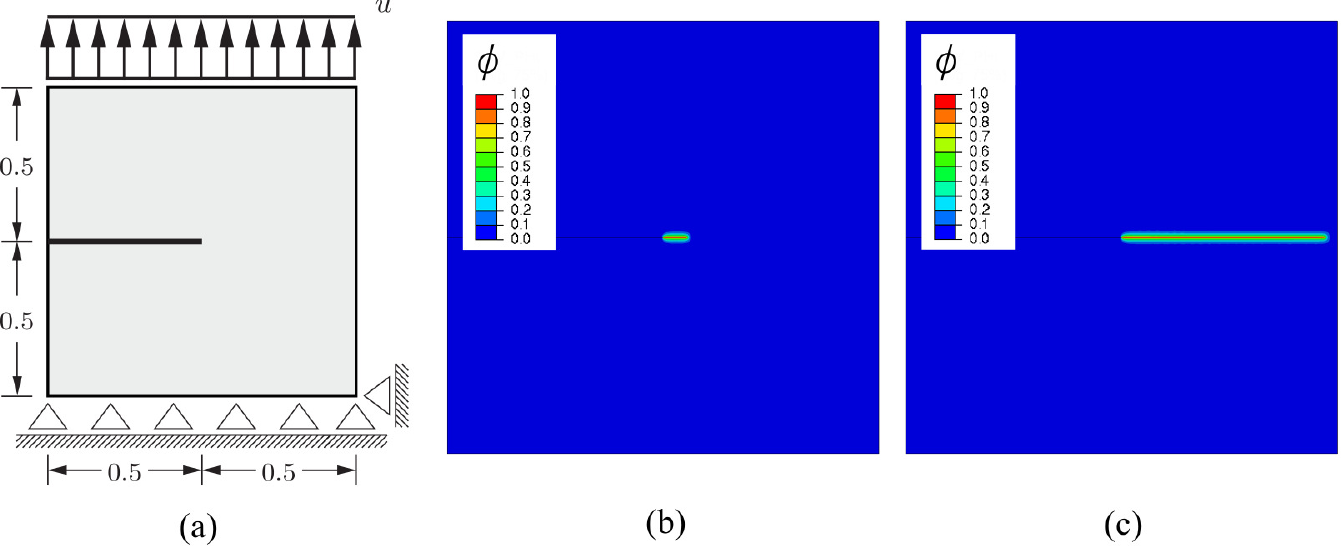}
\caption{Cracked square plate: (a) Loading configuration (with dimensions in mm) and phase field contours after (b) 80 and (c) 280 loading cycles.}
\label{fig:SENT_Geometry}
\end{figure}

\begin{figure}[H]
\centering
\includegraphics[width=0.7\linewidth]{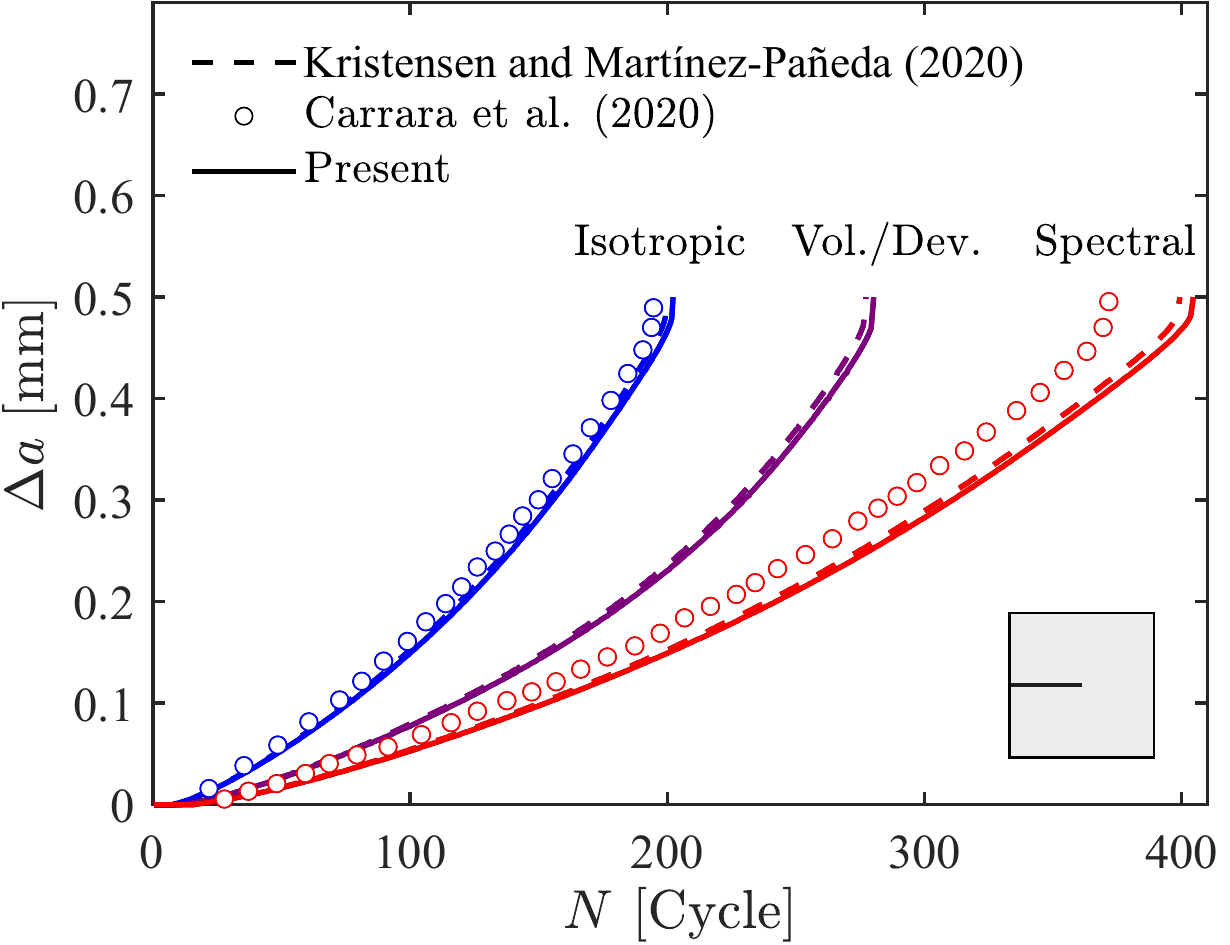}
\caption{Cracked square plate, validation in an inert environment: crack extension versus number of cycles and comparison with the results of Kristensen and Mart\'{\i}nez-Pa\~neda \cite{TAFM2020} and Carrara \textit{et al.} \cite{Carrara2020}.}
\label{fig:SENT_PsiDecomposition}
\end{figure}

\begin{figure}[H]
\centering
\includegraphics[width=0.7\linewidth]{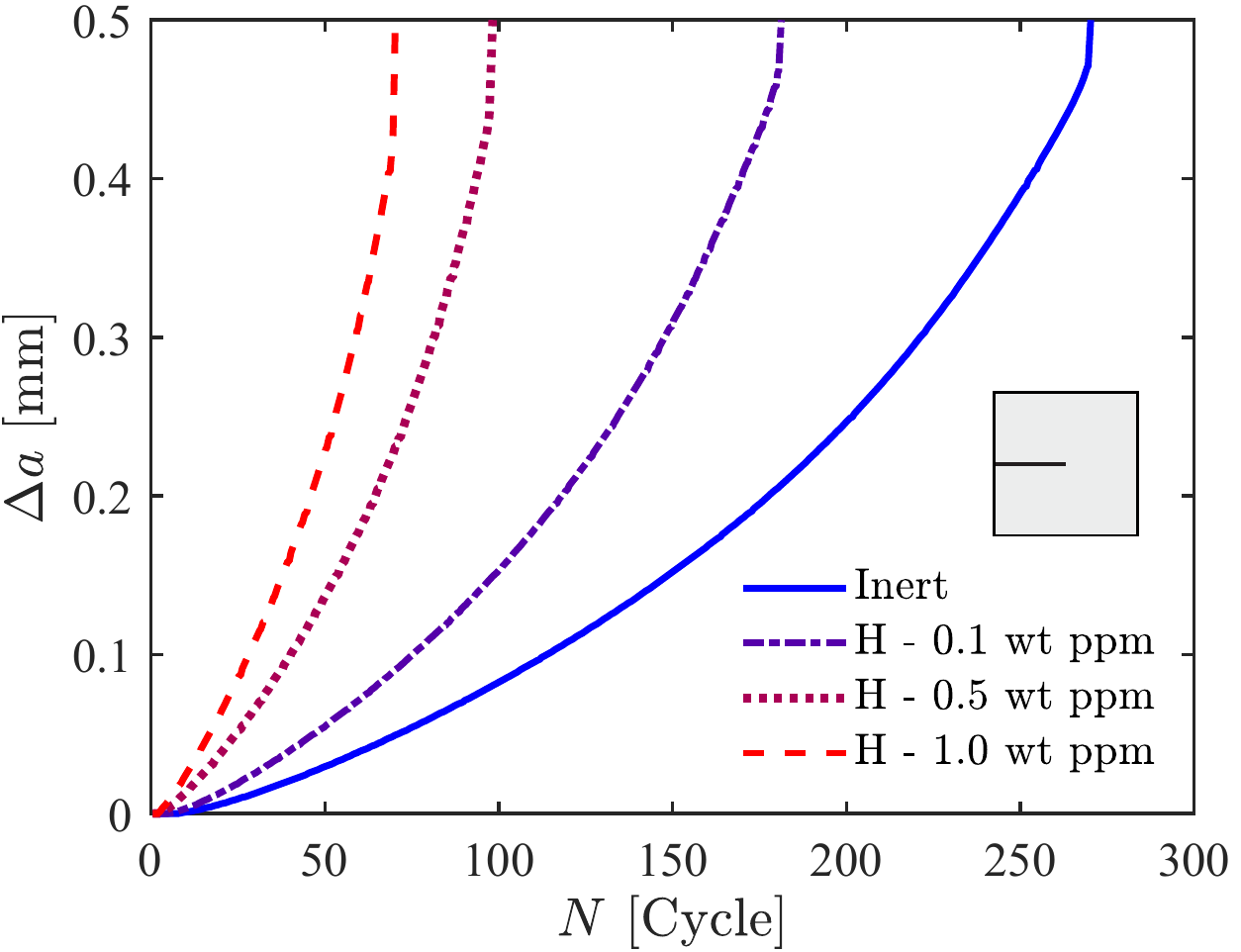}
\caption{Cracked square plate, influence of hydrogen: crack extension versus number of cycles for various hydrogen concentration levels.}
\label{fig:SENT_aVsN_h}
\end{figure}

The case of a square plate with an initial crack subjected to uniaxial tension has become a paradigmatic benchmark in the phase field fracture community. Loading conditions and sample dimensions (in mm) are illustrated in Fig. \ref{fig:SENT_Geometry}a. As in Refs. \cite{Carrara2020,TAFM2020}, material properties read $E=210$ GPa, $\nu=0.3$, $G_c=2.7$ kJ/m$^2$, $\ell=0.004$ mm and $\alpha_T=56.25$ MPa. The sample is discretised using 27,410 eight-node plane strain quadrilateral elements with reduced integration. The mesh is refined in the crack propagation region to ensure that the characteristic element length $h$ is sufficiently small to resolve the fracture process zone ($h < \ell/5.4$ \cite{CMAME2018}). The plate is subjected to a piece-wise linear cyclic remote displacement with a load frequency of $f=400\:\text{Hz}$, a zero mean value (i.e. a load ratio of $R=-1$) and a constant range of $\Delta u=4\times10^{-3}$ mm.
We proceed first to validate the model in the absence of hydrogen. The results obtained are shown in Fig. \ref{fig:SENT_PsiDecomposition} in terms of crack extension $\Delta a$ (in mm) versus the number of cycles $N$. The computations have been conducted for three choices of the strain energy density decomposition: no split, volumetric/deviatoric \cite{Amor2009} and spectral \cite{Miehe2010a}. A very good agreement is observed with the predictions of Carrara \textit{et al.} \cite{Carrara2020} and Kristensen and Mart\'{\i}nez-Pa\~neda \cite{TAFM2020}. The agreement is particularly good with the latter work, which uses a quasi-Newton monolithic implementation, while the work by Carrara \textit{et al.} \cite{Carrara2020} employs an energy-based criterion to ensure that the staggered solution scheme iterates until reaching the monolithic solution \cite{Ambati2015}. As discussed in the literature, higher fatigue crack growth rates are predicted if no tension-compression split is considered as both tension and compression loading cycles contribute to damage.
Subsequently, the cracked square plate is exposed to a hydrogenous environment at room temperature. We assume that the plate is made of an iron-based material with  diffusion coefficient $D=0.0127$ mm$^2$/s. Furthermore, it is assumed that the sample has been pre-charged and is exposed to a hydrogenous environment throughout the experiment. Accordingly, a uniform hydrogen distribution is assigned as an initial condition $C(t=0)=C_0=C_\mathrm{env} \,\,\, \forall \, x$ and a constant hydrogen concentration $C(t)=C_\mathrm{env}$ is prescribed at all the outer boundaries of the plate, including the crack faces\footnote{We note that, while a constant hydrogen concentration has been prescribed at the crack faces for simplicity, the use of generalised Neumann-type boundary conditions \cite{Turnbull1996,CS2020b} or $\sigma_H$-dependent Dirichlet boundary conditions \cite{DiLeo2013,IJHE2016,Diaz2016b} is more appropriate.}. The results obtained are shown in Fig. \ref{fig:SENT_aVsN_h} for three selected values of the environmental hydrogen concentration: 0.1, 0.5 and 1 wt ppm. The results reveal that the model correctly captures the trend expected: fatigue crack growth rates increase with increasing hydrogen content (see, e.g., \cite{Gangloff1990,Gangloff2012}).

\subsection{Boundary layer model} 
\label{sec:4_BoundaryLayer}

\begin{figure}[H]
\centering
\includegraphics[width=1\linewidth]{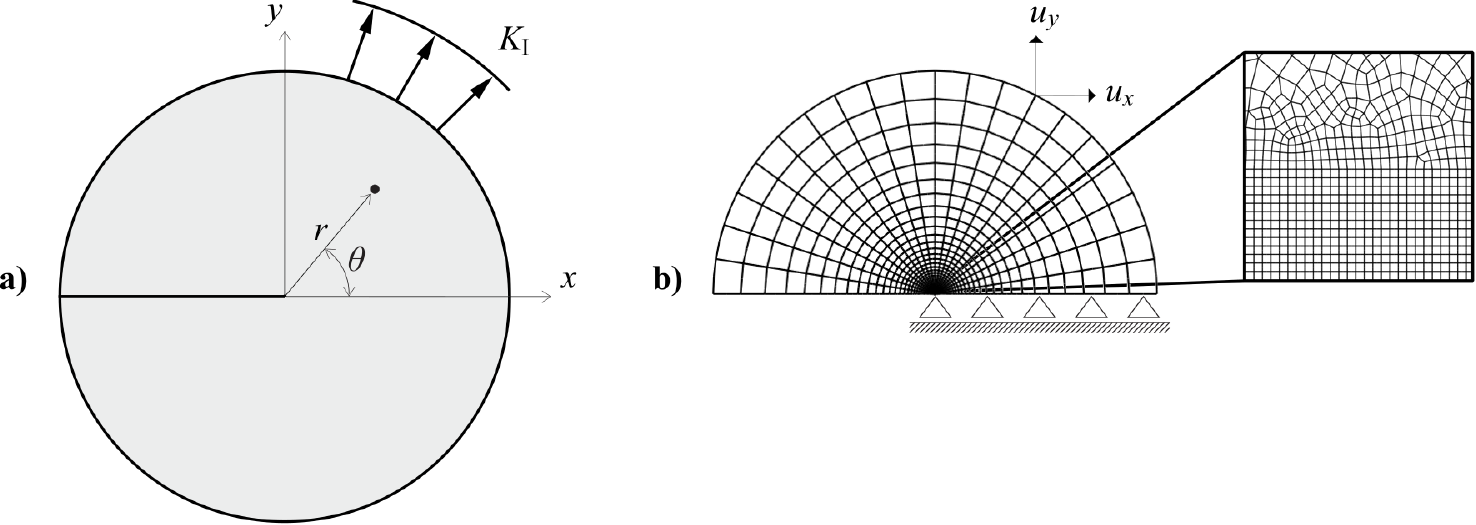}
\caption{Boundary layer model: (a) Geometry and boundary conditions, and (b) finite element mesh, including details of the mesh refinement ahead of the crack tip.}
\label{fig:BL-GeometryMesh}
\end{figure}

Next, we gain insight into hydrogen-assisted fatigue under small scale yielding conditions. A boundary layer model is used to prescribe a remote $K_{\mathrm{I}}$ field in a circular region of a body containing a sharp crack. As shown in Fig. \ref{fig:BL-GeometryMesh}, only the upper half of the domain is considered due to its symmetry. The remote, elastic $K_{\mathrm{I}}$ field is applied by prescribing the displacements of the nodes in the outer region following the Williams \cite{Williams1957} expansion. Thus, for a polar coordinate system ($r,\theta$) centered at the crack tip, the horizontal and vertical displacements respectively read
\begin{equation}
\begin{aligned}
&u_x(r, \theta)=K_{\mathrm{I}} \frac{1+\nu}{E} \sqrt{\frac{r}{2 \pi}} \cos{\frac{\theta}{2}}\big[3-4 \nu-\cos{\theta}\big]\\[3mm]
&u_y(r, \theta)=K_{\mathrm{I}} \frac{1+\nu}{E} \sqrt{\frac{r}{2 \pi}} \sin{\frac{\theta}{2}}\big[3-4 \nu-\cos{\theta}\big]
\end{aligned}
\end{equation}

Cyclic loading conditions are attained by defining the applied stress intensity factor as the following sinusoidal function,
\begin{equation}
    K_\mathrm{I}=K_{\text{m}}+ \dfrac{\Delta K}{2}\sin{2\pi f\,t}\,, \hspace{10mm}\text{with}\hspace{5mm} K_{\text{m}}=\dfrac{\Delta K}{2}+\dfrac{R\,\Delta K}{1-R}
\end{equation}
\noindent where $f$ denotes the load frequency, $t$ the test time, $K_{\text{m}}$ the load mean value, $\Delta K=K_{\text{max}}-K_{\text{min}}$ the load range, and $R=K_{\text{min}}/K_{\text{max}}$ the load ratio. To capture the loading history with fidelity, each cycle is divided into at least 20 computational time increments. The circular domain is discretised using 4,572 quadratic plane strain quadrilateral elements with reduced integration and, as shown in Fig. \ref{fig:BL-GeometryMesh}b, the mesh is refined along the crack propagation region.
Consider first the case of a stationary crack in a solid with Young's modulus $E=210$ GPa, Poisson's ratio $\nu=0.3$ and diffusion coefficient \(D=0.0127\:\text{mm}^2/\text{s}\). The sample is assumed to be pre-charged with a uniform concentration of $C(t=0)=C_0=0.5$ wt ppm. The load range is chosen to be $\Delta K=1\:\text{MPa}\sqrt{\text{m}}$, the load frequency equals $f=1\:\text{Hz}$, and the load ratio is $R=0$. The evolution of the crack tip hydrogen distribution as a function of time $t$ can be quantified by the following dimensionless groups, as dictated by the Buckingham $\Pi$ theorem,  
\begin{equation}
    \dfrac{C}{C_0}=\mathcal{F}\left(\dfrac{f L_0^2}{D},\quad \dfrac{t D}{L^2_0} , \quad \dfrac{E\bar{V}_{H}}{\mathcal{R} T}\right)
\end{equation}
\noindent where $L_0=(K_{\text{m}}/E)^2$ is a length parameter that results from the dimensional analysis and provides a measure of the gradients close to the crack tip. The first two dimensionless groups quantify the competing influence of test and diffusion times, which are denoted as the normalised frequency $\bar{f}=\dfrac{f L_0^2}{D}$ and the normalised time $\bar{t}=\dfrac{t D}{L^2_0}$, respectively.\\

Hydrogen diffusion is (partially) driven by gradients of hydrostatic stress, see Eq. (\ref{eq:chemicalpotential}), such that hydrogen atoms will accumulate in areas with high volumetric strains. Under steady state conditions, the hydrogen concentration is given as,
\begin{equation}\label{eq:SteadyState}
    C = C_0  \exp{\frac{\bar{V}_H \sigma_H}{\mathcal{R}T}} \, .
\end{equation}

\begin{figure}[H]
\centering
\includegraphics[width=0.7\linewidth]{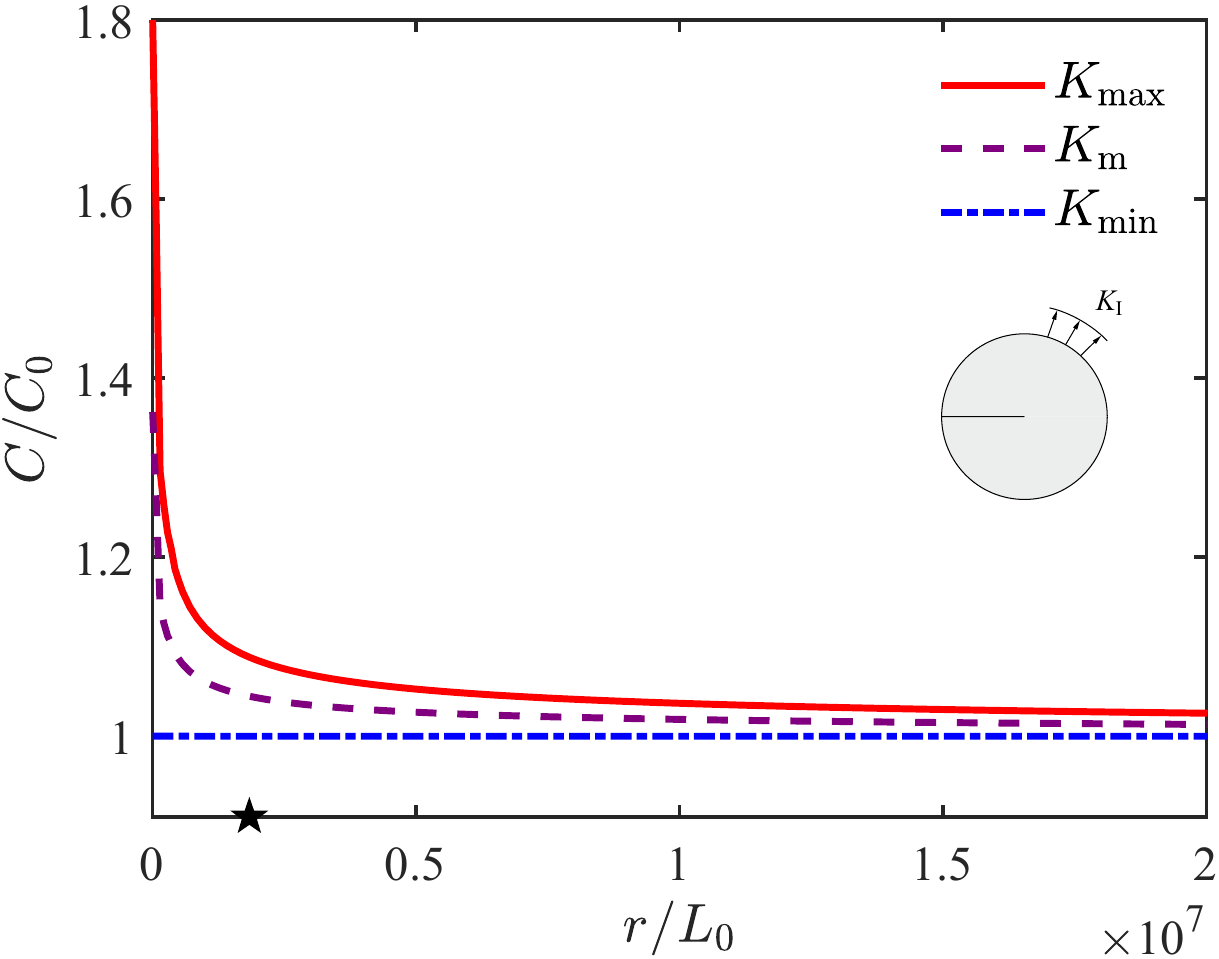}
\caption{Boundary layer model: Hydrogen concentration ahead of a stationary crack tip for three stages of the first load cycle. The results have been obtained under steady state conditions and with load ratio $R=0$.}
\label{fig:1_CL-curve}
\end{figure}

Accordingly, the hydrogen distribution ahead of the crack will vary during the loading cycle. Fig. \ref{fig:1_CL-curve} shows the results obtained at the maximum $K_{\text{max}}$, mean $K_{\text{m}}$ and minimum $K_{\text{min}}=0$ stages of the first load cycle, for a sufficiently low frequency such that conditions resemble those of steady state. In agreement with expectations, the hydrogen concentration increases with the applied load, reaching its maximum value in the vicinity of the crack tip (where $\sigma_H$ is highest), and remains constant for a zero value of the hydrostatic stress at $K_\mathrm{min}=0$ ($R=0$).

\begin{figure}[H]
\captionsetup{width=0.99\textwidth}
    \centering
    \begin{minipage}{0.99\textwidth}
    \centering
    \includegraphics[width=0.99\linewidth]{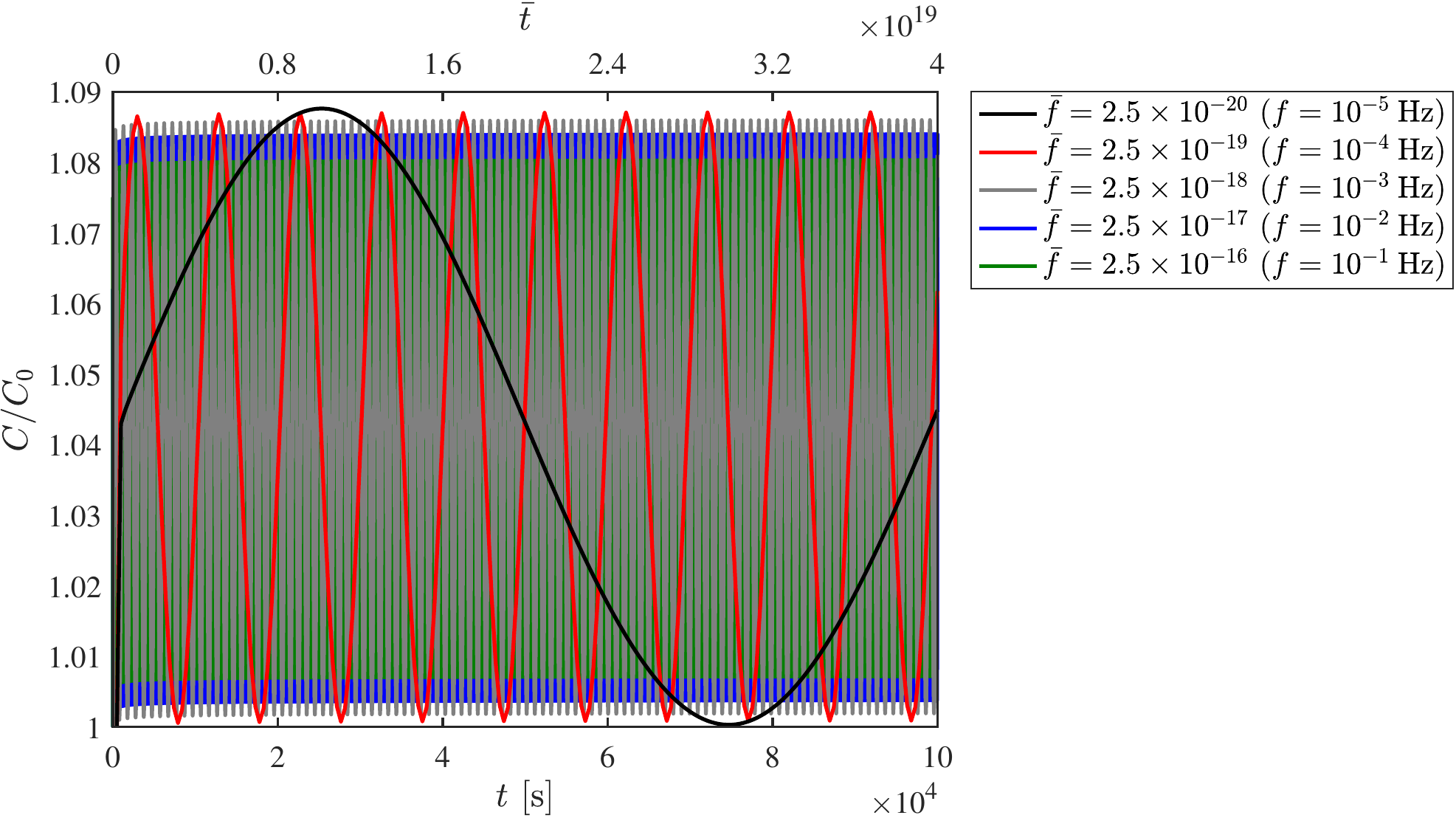}
    \end{minipage}
    \\[3mm]
    \begin{minipage}{0.99\textwidth}
    \centering
    \includegraphics[width=0.99\linewidth]{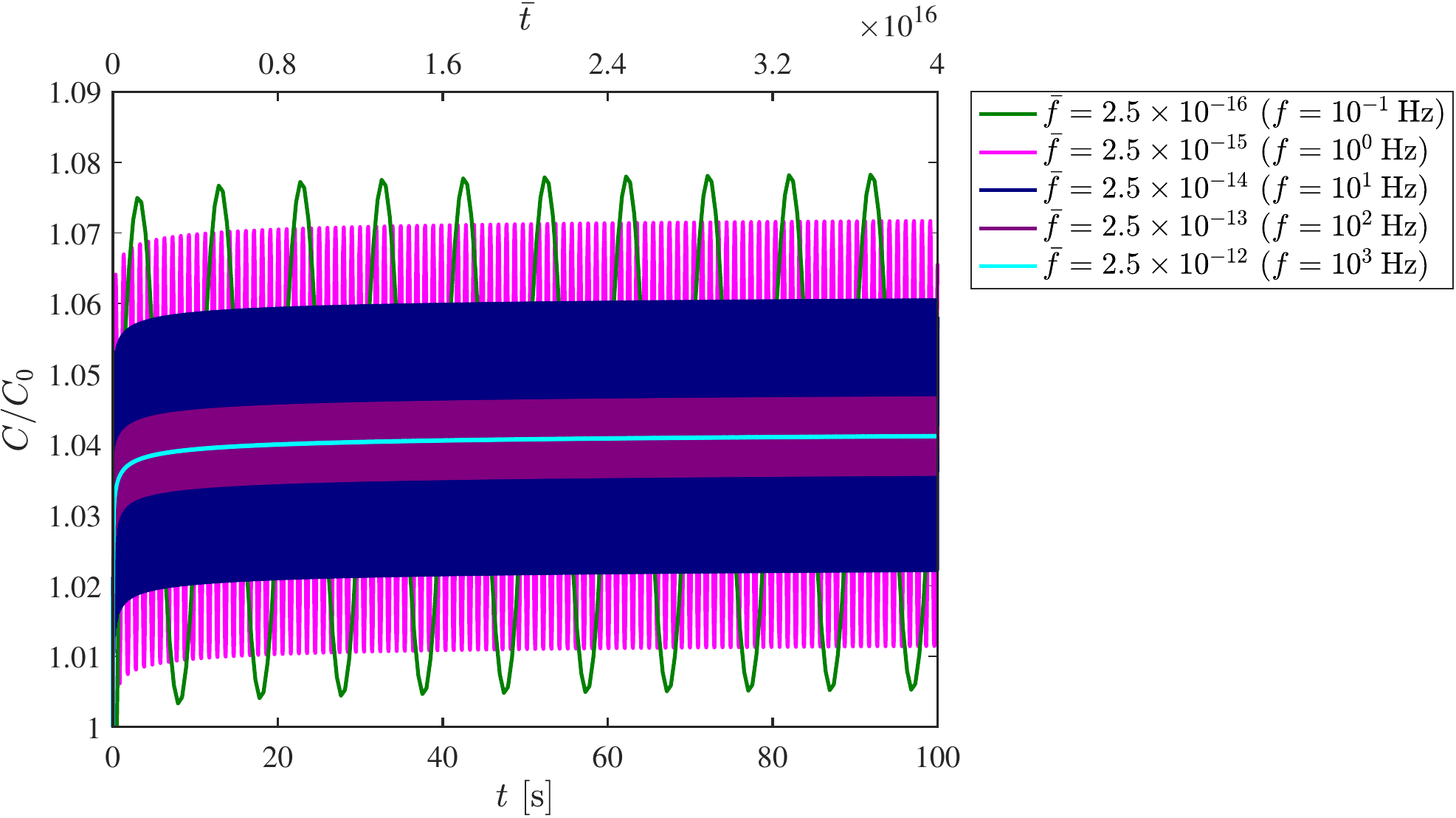}
    \end{minipage}
    \captionof{figure}{Boundary layer model: Variation in time of the hydrogen concentration at a point ahead of a stationary crack tip for various loading frequencies and load ratio $R=0$.}
    \label{fig:1_CL_time}
\end{figure}

Let us now consider the more common case of transient conditions and investigate the competing role of the loading frequency and diffusion time. Fig. \ref{fig:1_CL_time} illustrates the variation in time of the hydrogen concentration near the crack tip, at a point located at $r/L_0\approx0.2\times10^7$, as denoted by a star in Fig. \ref{fig:1_CL-curve}. The results reveal that, irrespectively of the test duration, the maximum hydrogen content that can be attained ahead of the crack tip is sensitive to the loading frequency. If the diffusivity of hydrogen is sufficiently large relative to the time required to complete one cycle (low $f$), the amplitude of the hydrogen concentration follows that of the hydrostatic stress, as in the steady state case - see Eq. (\ref{eq:SteadyState}). Contrarily, for high loading frequencies, unloading begins before the hydrogen distribution reaches the steady state solution (\ref{eq:SteadyState}) and consequently the maximum value of $C$ reached during the experiment is smaller than that of lower frequencies. It can be seen that, for the highest frequency ($f=10^{3}\,\mathrm{Hz}$) the hydrogen concentration does not oscillate and flattens out towards a constant value that is roughly 5\% lower than the maximum concentration attained at low loading frequencies (for the material properties and distance ahead of the crack here considered). Recall that the relevant non-dimensional group $\bar{f}=f L_0^2/D$ involves the material diffusion coefficient. It follows that the present results could support the use of \emph{beneficial} traps, which lower the material diffusivity but are not involved in the fracture process, as a viable strategy for designing materials resistant to hydrogen-assisted fatigue. 

We proceed to investigate the influence of the diffusion time-frequency interplay on fatigue crack growth rates. The phase field fatigue model outlined in Section \ref{Sec:Theory} is used, with material properties \(G_{c}=2.7 \: \text{kJ}/\text{m}^2\) and $\ell=0.0048$ mm. A reference stress intensity factor, in the absence of hydrogen, is defined as,
\begin{equation}
    K_0=\sqrt{\dfrac{G_c E}{\left(1-\nu^2\right)}}
\end{equation}

\noindent and a fracture process zone length $L_f$, can be defined as \cite{Tanne2018,PTRSA2021}:
\begin{equation}
    L_f = \frac{G_c \left( 1 - \nu^2 \right)}{E}
\end{equation}

\begin{figure}[H]
\centering
\includegraphics[width=0.7\linewidth]{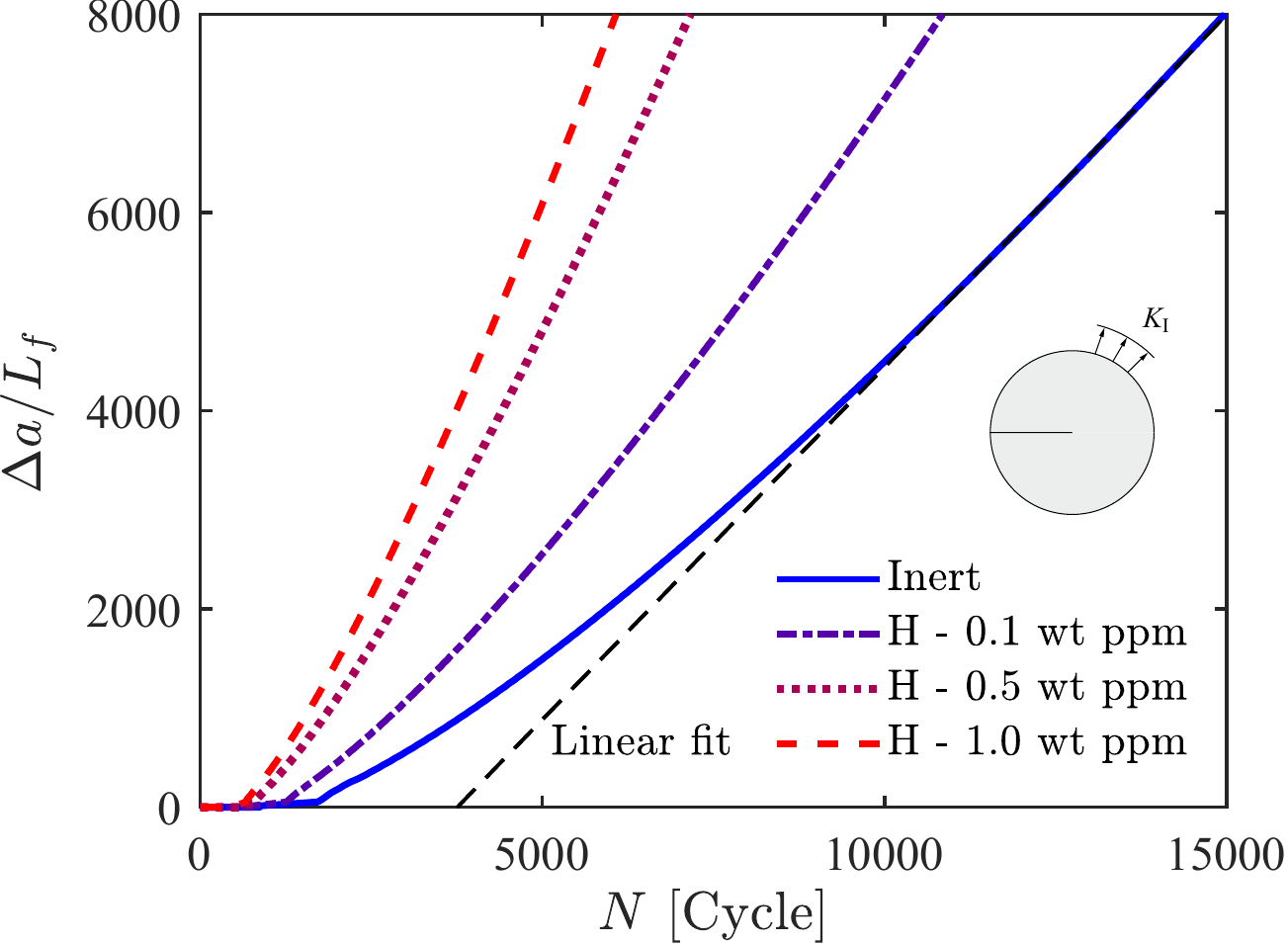}
\caption{Boundary layer model: Crack extension versus the number of cycles for different hydrogen concentrations. Results have been obtained for $\Delta K/K_0=0.08$, under a load ratio of $R=0.1$ and load frequency $f=1\:\text{Hz}$.}
\label{fig:2_aVsN}
\end{figure}

Fig. \ref{fig:2_aVsN} shows the results obtained in terms of (normalised) crack extension versus number of cycles, as a function of the environmental hydrogen concentration $C_\mathrm{env}$. These computations have been conducted for a pre-charged solid ($C(t=0)=C_\mathrm{env}$) that is exposed to a hydrogenous environment during the test ($C(t)=C_\mathrm{env}$ at the boundaries). The load range equals $\Delta K/K_0=0.08$, while the load frequency and ratio equal $f=1\:\text{Hz}$ and $R=0.1$, respectively. The results shown in Fig. \ref{fig:2_aVsN} reveal that the model is able to capture the expected trends - for a given number of cycles, the higher the hydrogen concentration, the larger the crack extension. As depicted in Fig. \ref{fig:2_aVsN}, a linear fit can be applied to the linear part of the curve to derive the slope (crack growth rates).

\begin{figure}[H]
\centering
\includegraphics[width=0.7\linewidth]{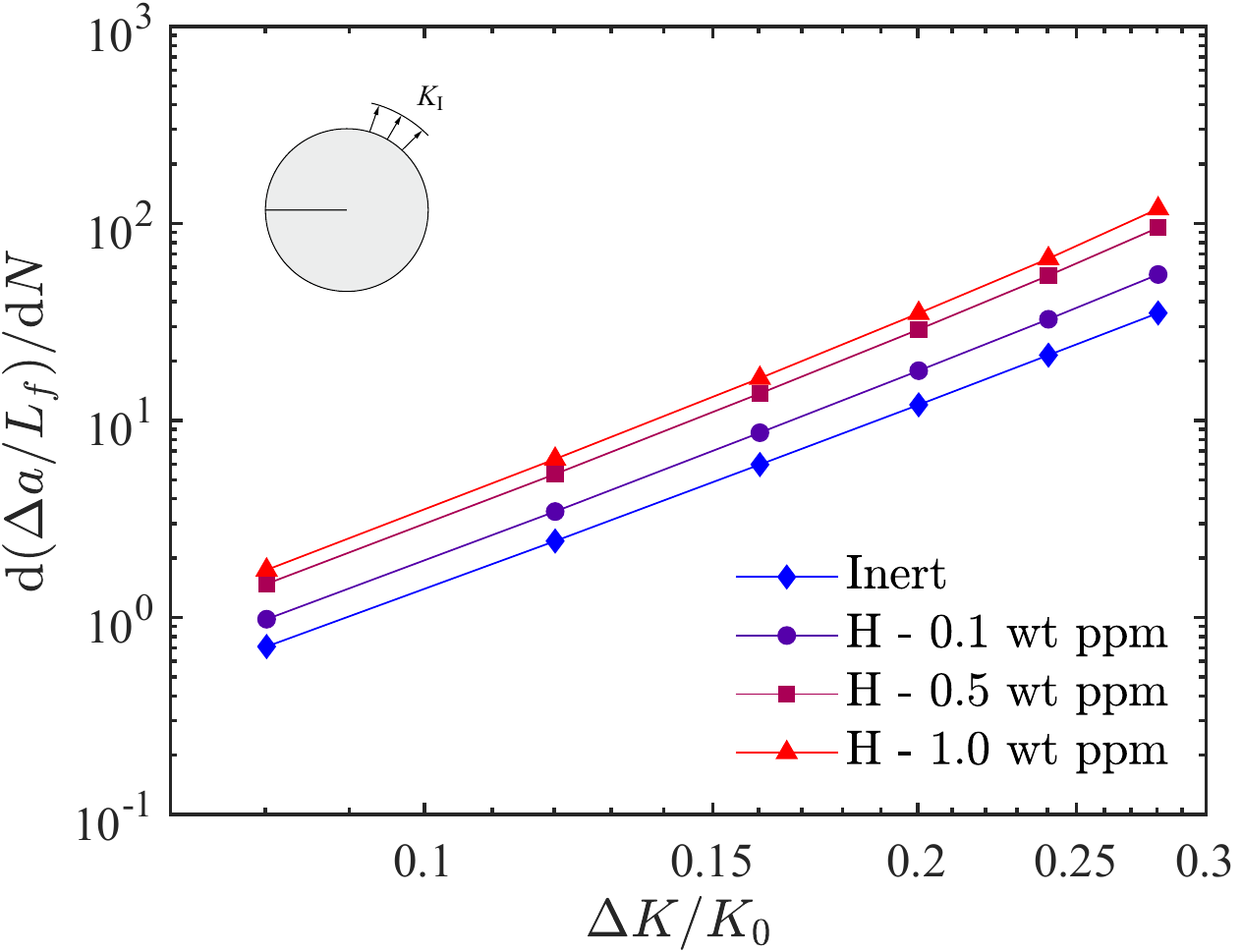}
\caption{Boundary layer model, Paris law behaviour: Fatigue crack growth rate versus load range for different hydrogen concentrations. Results have been obtained for a load ratio of $R=0.1$ and load frequency $f=1\:\text{Hz}$.}
\label{fig:2_Paris}
\end{figure}

The fatigue crack growth rates obtained for different $\Delta K$ and hydrogen concentrations are shown in Fig. \ref{fig:2_Paris}, using a log-log plot. The computed curves behave linearly in the so-called Paris regime, where cracks propagate stably, as expected. By applying the well-known Paris equation $\mathrm{d}a/\mathrm{d}N= \mathcal{C} \Delta K^m$, one can readily observe that $\mathcal{C}$ increases with the hydrogen content, in agreement with the experimental trends. On the other hand, results yield a Paris exponent that appears to be less sensitive to the environment, with a magnitude ($m\approx 3.2$) that is within the range reported for metals in inert environments \cite{Anderson2005}. The present framework is capable of providing as an output (not input) the Paris law behaviour, enabling the prediction of the role of hydrogen in accelerating sub-critical crack growth rates. 

\begin{figure}[H]
\centering
\includegraphics[width=0.7\linewidth]{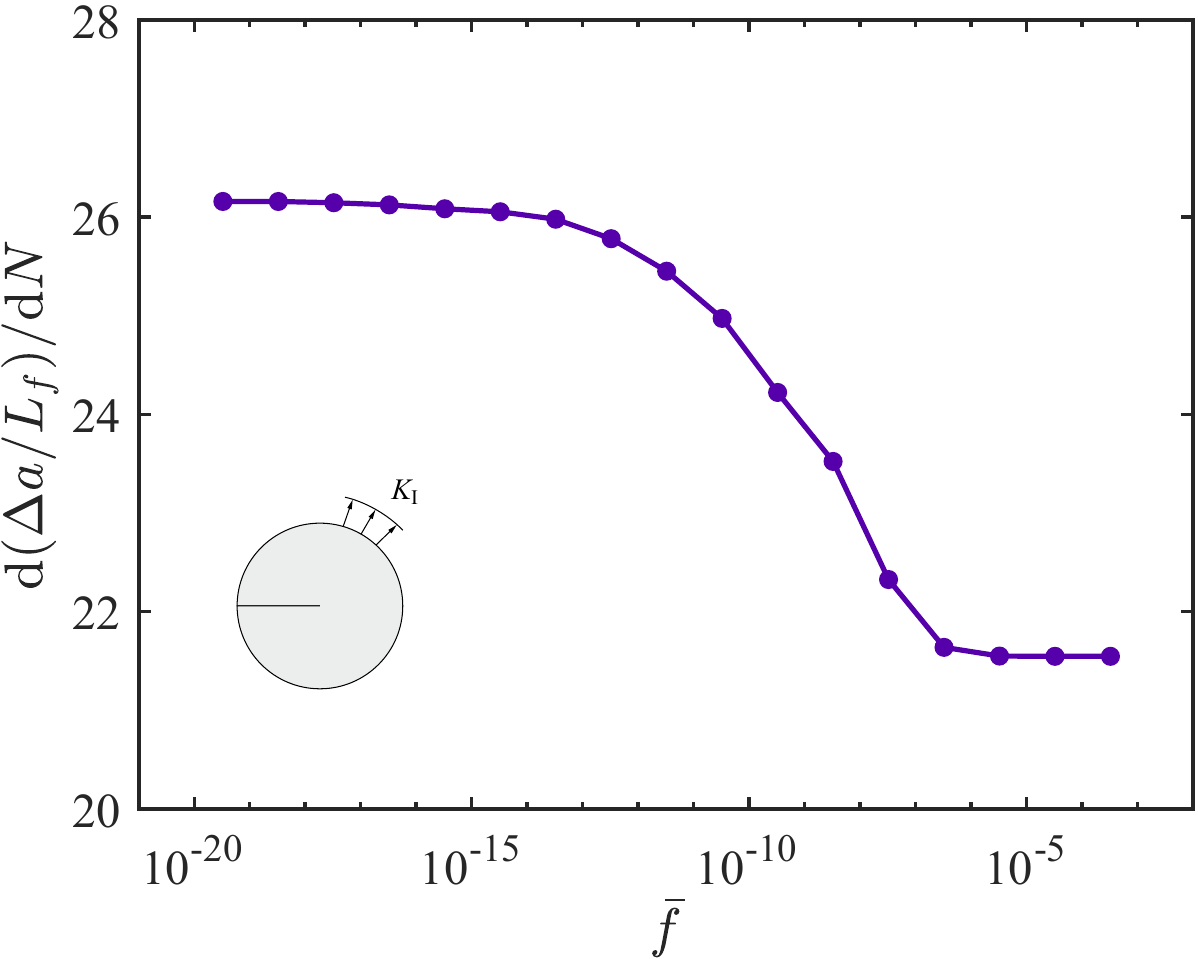}
\caption{Boundary layer model, mapping frequency regimes: fatigue crack growth rate versus normalised frequency $\bar{f}=fL_0^2/D$. Results have been obtained for $\Delta K/K_0=0.24$, under a load ratio of $R=0$ and a hydrogen concentration of $C_0=C_\mathrm{env}=0.1\:\text{wt ppm}$.}
\label{fig:2_dadN_vs_freq}
\end{figure}

Finally, Fig. \ref{fig:2_dadN_vs_freq} illustrates the sensitivity of fatigue crack growth rates to the loading frequency. Here, we consider a pre-charged sample with $C_0=0.1$ wt ppm exposed to a load amplitude of $\Delta K / K_0=0.24$ and a load ratio of $R=0$. It is shown that the model captures another widely observed experimental trend; the fatigue behaviour of metals in the presence of hydrogen varies between two limiting cases: (i) fast tests (high $f$), where hydrogen does not have enough time to diffuse to the fracture process zone and the susceptibility to embrittlement diminishes, and (ii) slow tests (low $f$), where hydrogen atoms have sufficient time to accumulate in areas of high $\sigma_H$, magnifying embrittlement. The model readily captures the transition between these two limiting regimes. 

\subsection{Notched cylindrical bar}
\label{sec:4_NotchedBar}

\begin{figure}[H]
\centering
\includegraphics[width=0.8\linewidth]{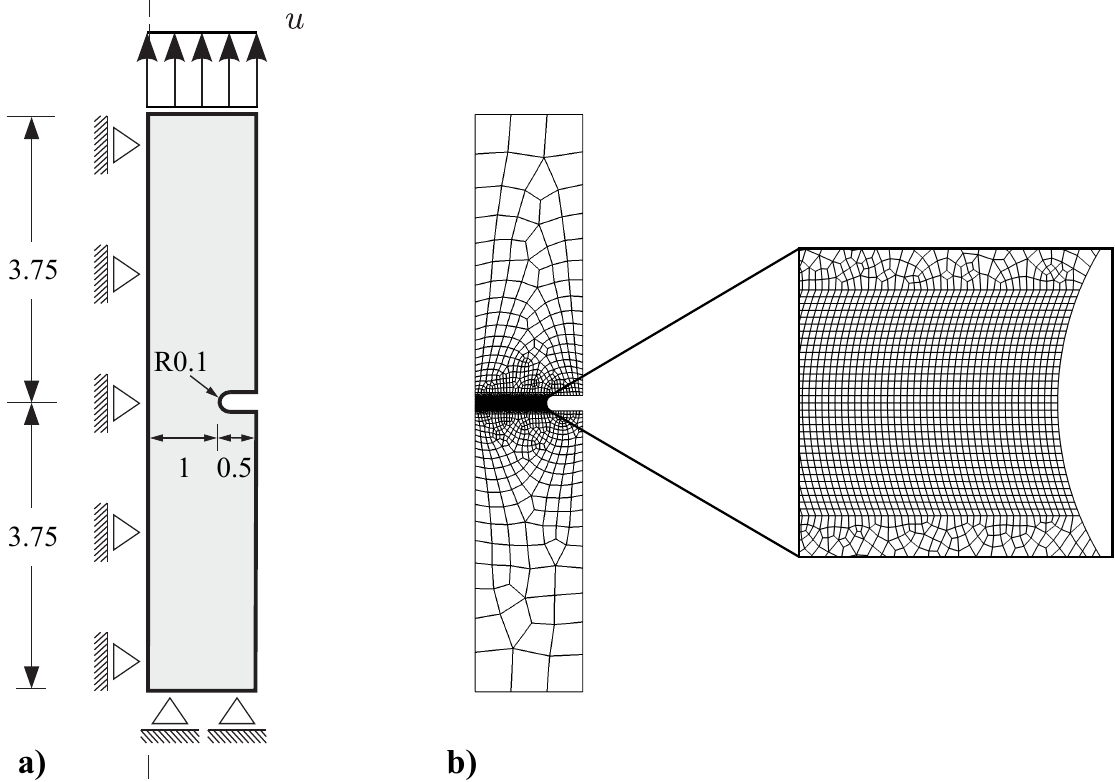}
\caption{Notched cylindrical bar: (a) geometry (with dimensions in mm) and boundary conditions, and (b) finite element mesh, including a detailed view of the mesh ahead of the notch tip.}
\label{fig:Bar_Geometry}
\end{figure}

Fatigue crack growth in samples containing non-sharp defects is subsequently investigated. Consider a cylindrical bar with a notch on its surface, as sketched in Fig. \ref{fig:Bar_Geometry}a. Axisymmetric conditions are exploited to model one planar section of the sample only. The finite element model contains 17,003 quadratic axisymmetric quadrilateral elements with reduced integration, with the mesh being refined ahead of the notch tip, where the characteristic element size is 6 times smaller than the phase field length scale $\ell$ (see Fig. \ref{fig:Bar_Geometry}b). The assumed material properties read $E=210$ GPa, $\nu=0.3$, $G_c=64$ kJ/m$^2$, $\ell=0.015$ mm, $D=0.0127$ mm$^2$/s, and $\alpha_T=355.56$ MPa. The bar is pre-charged and subsequently loaded in the same environment such that all the outer boundaries of the bar, including the notch faces, are in contact with the environment during the entire numerical experiment. Three environments are considered, corresponding to hydrogen concentrations of 0.1, 0.5 and 1 wt ppm. Cyclic loading is prescribed by subjecting the bar to a piece-wise linear remote displacement with a load frequency of $f=1\:\text{Hz}$ and a load ratio of $R=0$. 

\begin{figure}[H]
\centering
\includegraphics[width=0.7\linewidth]{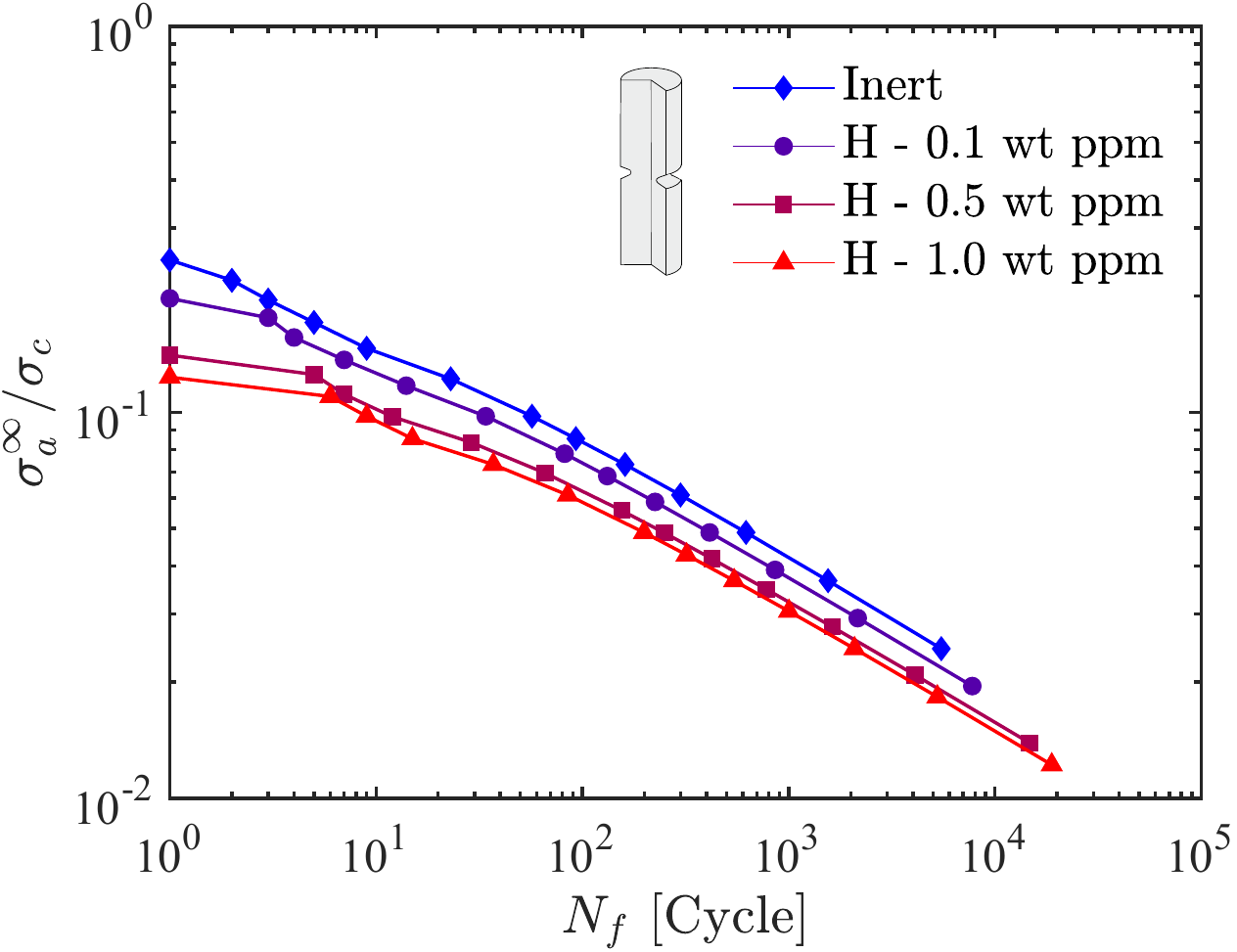}
\caption{Notched cylindrical bar, \emph{Virtual} S-N curves: alternating remote stress versus number of cycles to failure for different hydrogen concentrations. The stress concentration factor equals $K_t=3.354$.}
\label{fig:3_SN_H2}
\end{figure}

The results obtained are shown in Fig. \ref{fig:3_SN_H2}, in terms of the remote stress amplitude versus the number of cycles to failure, also known as S-N curves. The stress amplitude is normalised by the material strength, as given by (\ref{eq:Sc}). For a given hydrogen concentration, shorter fatigue lives are observed as the hydrogen content is increased. In all cases, the number of cycles to failure increases with decreasing stress amplitude, and the slope of the S-N curve appears to be rather insensitive to the hydrogen content. 

\begin{figure}[H]
\begin{subfigure}[t]{0.49\textwidth}
        \centering
    \includegraphics[width=0.85\linewidth,trim={0 0 0 0},clip]{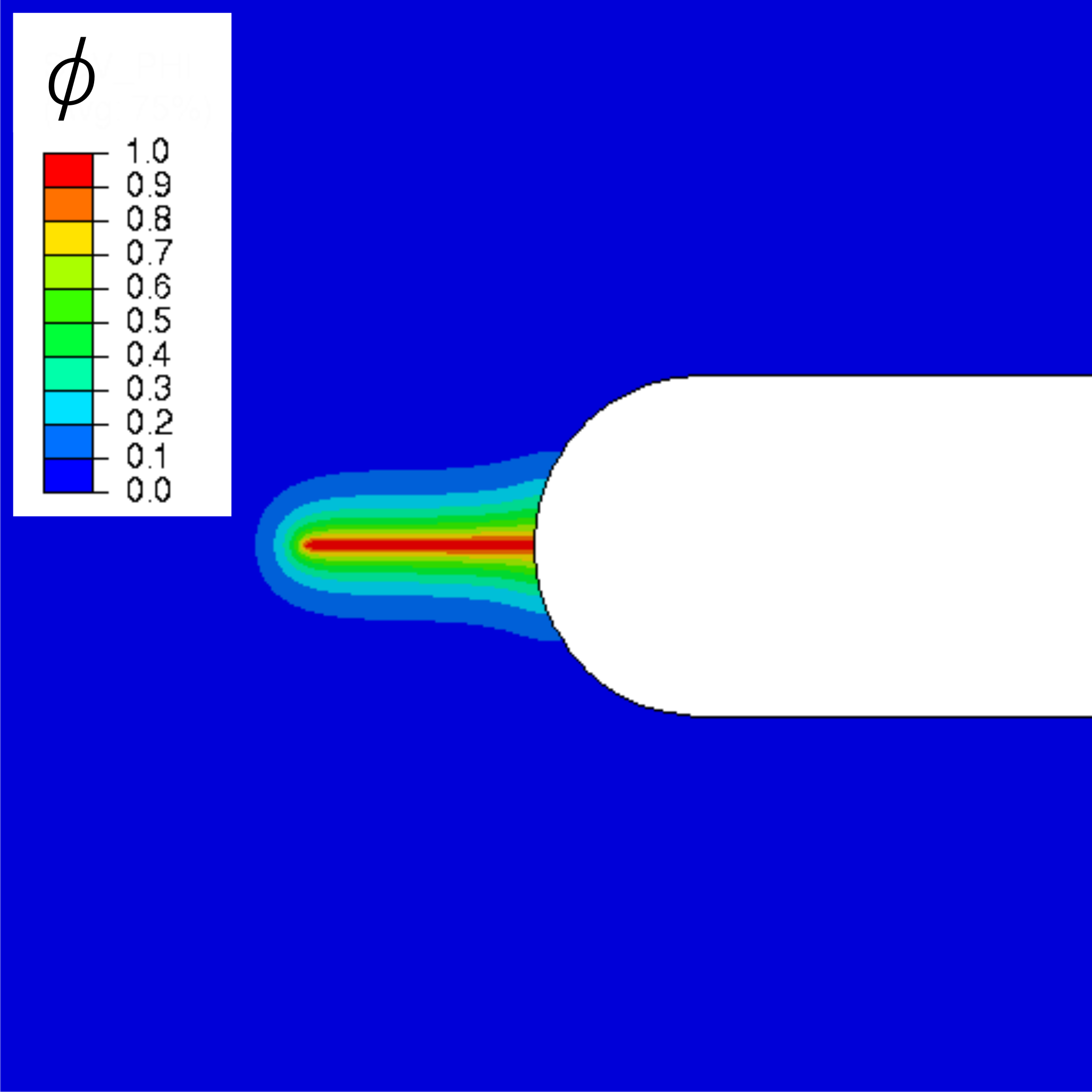}
    \caption{}
    \label{}
    \end{subfigure}
    \begin{subfigure}[t]{0.49\textwidth}
        \centering
    \includegraphics[width=0.85\linewidth,trim={0 0 0 0},clip]{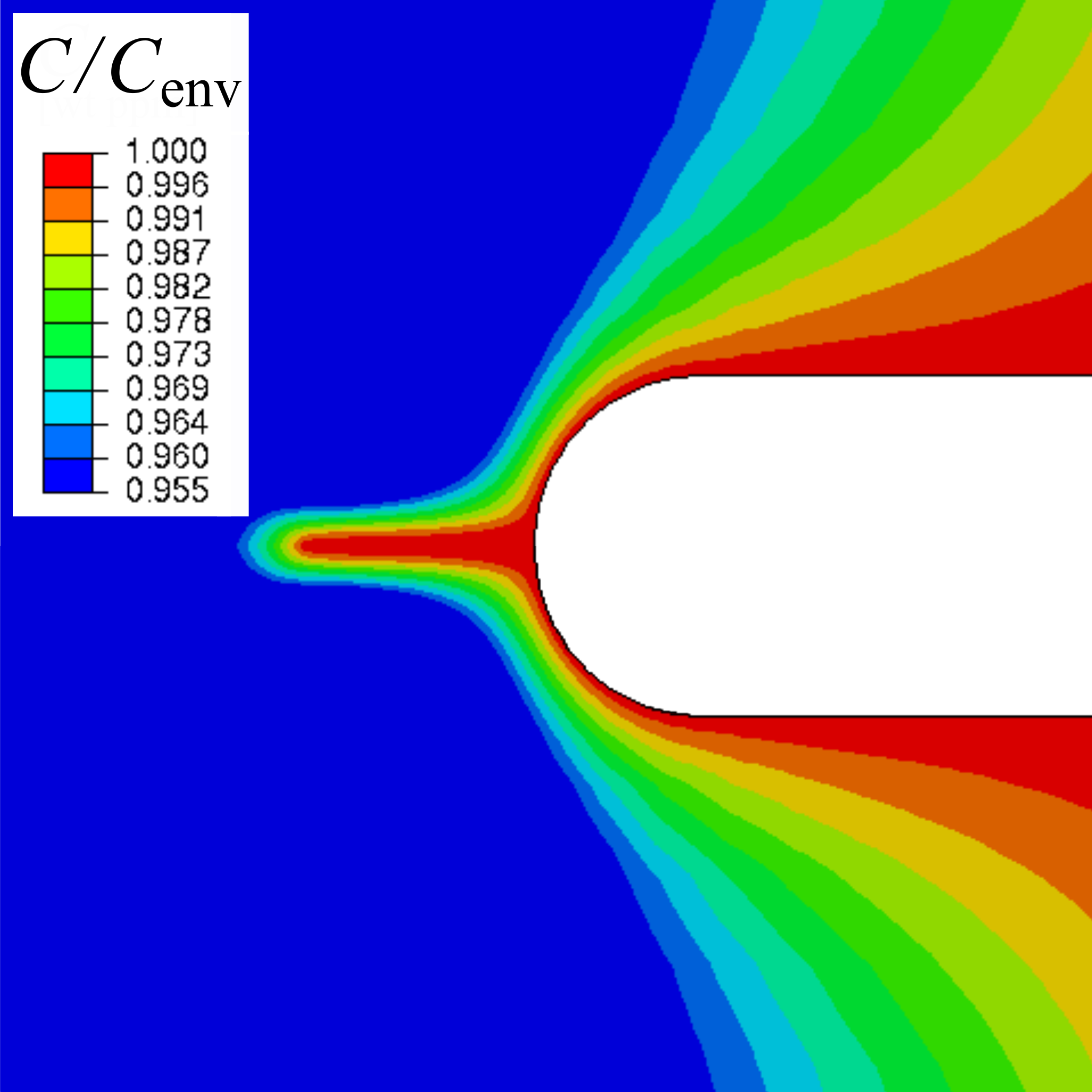}
    \caption{}
    \label{}    
    \end{subfigure}
    \caption{Notched cylindrical bar, influence of the \emph{moving} chemical boundary conditions: contours of the phase field $\phi$ (a) and hydrogen concentration (b). Results have been obtained for $C_{\mathrm{env}}=1\:\text{wt ppm}$ after $700$ cycles and are plotted at $u=u_\mathrm{min}=0$.}
    \label{fig:3_contours}
\end{figure}

Accurate fatigue crack growth predictions in harmful environments require suitable boundary conditions. As mentioned in Section \ref{Sec:Theory}, we adopt a penalty approach to implicitly enforce \emph{moving} chemical boundary conditions, so as to capture how the newly created crack surfaces are promptly exposed to the environment. This is illustrated in Fig. \ref{fig:3_contours} by means of phase field and hydrogen concentration contours; as the crack grows, the concentration in the damaged regions equals $C_{\text{env}}$. Note that the contours correspond to $\sigma^\infty=\sigma_{\text{min}}=0$, and as a result there is no effect of $\sigma_H$ on the hydrogen concentration. 
\subsection{Comparison with experimental S-N curves}
\label{sec:4_ExperimentalComparison}

We conclude the results section by comparing model predictions with S-N curves obtained from uniaxial tension-compression fatigue experiments on smooth samples. The tests were carried out by Matsunaga et al. \cite{Matsunaga2015} on two types of steels, a Cr-Mo steel (JIS-SCM435) with tensile strength of 840 MPa and a carbon steel (JIS-SM490B) with tensile strength of 530 MPa. The experiments were carried out in laboratory air and in $\SI{115}{MPa}$ hydrogen gas under constant stress amplitudes at a stress ratio of $R=-1$ and a test frequency of $f=\SI{1}{Hz}$. As it is common with steels, both materials are assumed to have a Young's modulus of $E=210$ GPa and a Poisson's ratio of $\nu=0.3$. The toughness is assumed to be equal to $G_c=60$ kJ/m$^2$ and $G_c=27$ kJ/m$^2$ for JIS-SCM435 and JIS-SM490B, respectively, based on fracture toughness measurements reported in Refs. \cite{Matsumoto2017,Ogawa2017}. The boundary value problem can be solved in a semi-analytical fashion, by considering the homogeneous solution to (\ref{eq:StrongFormEq}). A piece-wise cyclic linear variation of the remote stress is assumed. Under 1D conditions, the length scale and the strength are related via (\ref{eq:Sc}), and this relation renders magnitudes of $\ell=1.88$ mm and $\ell=2.13$ mm for JIS-SCM435 and JIS-SM490B, respectively. The logarithmic fatigue degradation function (\ref{eq:FdegLog}) is used, together with the spectral tension-compression split \cite{Miehe2010a}. The fatigue parameters $\alpha_T$ and $\kappa$ are chosen so as to provide the best fit to the experiments in air; the magnitudes of $\alpha_T=24$ MPa and $\kappa=0.15$ provided the best fit to both JIS-SCM435 and JIS-SM490B data. Then, the fatigue response of samples exposed to hydrogen can be estimated by relating the H$_2$ pressure with the hydrogen concentration. The latter can be given as a function of the solubility $S$ and the fugacity $f_{\text{H}_2}$ by means of Sievert's law: 
\begin{equation}
    C=S\sqrt{f_{\text{H}_2}} \hspace{10mm} \text{with}\hspace{5mm} S=S_0\exp{\dfrac{-E_s}{\mathcal{R} T}} \, ,
\end{equation}

\noindent where $E_s$ is an activation energy. For JIS-SCM435 and JIS-SM490B, the magnitudes of $S_0$ and $E_s$ are taken from Ref. \cite{SanMarchi2012} by considering the data reported for similar steels (AISI 4130 and AISI 1020, respectively); namely: $E_s=\SI{27.2}{kJ/mol}$, $S_0=\SI{102}{mol/m^3\sqrt{MPa}}$ (JIS-SCM435) and $E_s=\SI{23.54}{kJ/mol}$, $S_0=\SI{159}{mol/m^3\sqrt{MPa}}$ (JIS-SM490B). Assuming that the Abel–Noble equation is appropriate, the fugacity can be related to the hydrogen pressure $p$ as follows,
\begin{equation}
    f_{\text{H}_2} = p \exp{\frac{p b}{\mathcal{R} T}} \,
\end{equation}

\noindent where the Abel-Noble parameter is taken to be $b=15.84$ cm$^3$/mol, rendering $f_{\text{H}_2}=242.9$ MPa, and hydrogen concentrations of 0.00577 wt ppm (JIS-SCM435) and 0.04042 wt ppm (JIS-SM490B). The solubility dependence on the hydrostatic stress should also be accounted for; thus, we scale the hydrogen concentration according to (\ref{eq:SteadyState}) to determine the final magnitude of hydrogen uptake. 

\begin{figure}[H]
\begin{subfigure}[t]{0.49\textwidth}
        \centering
    \includegraphics[width=1\linewidth,trim={0 0 0 0},clip]{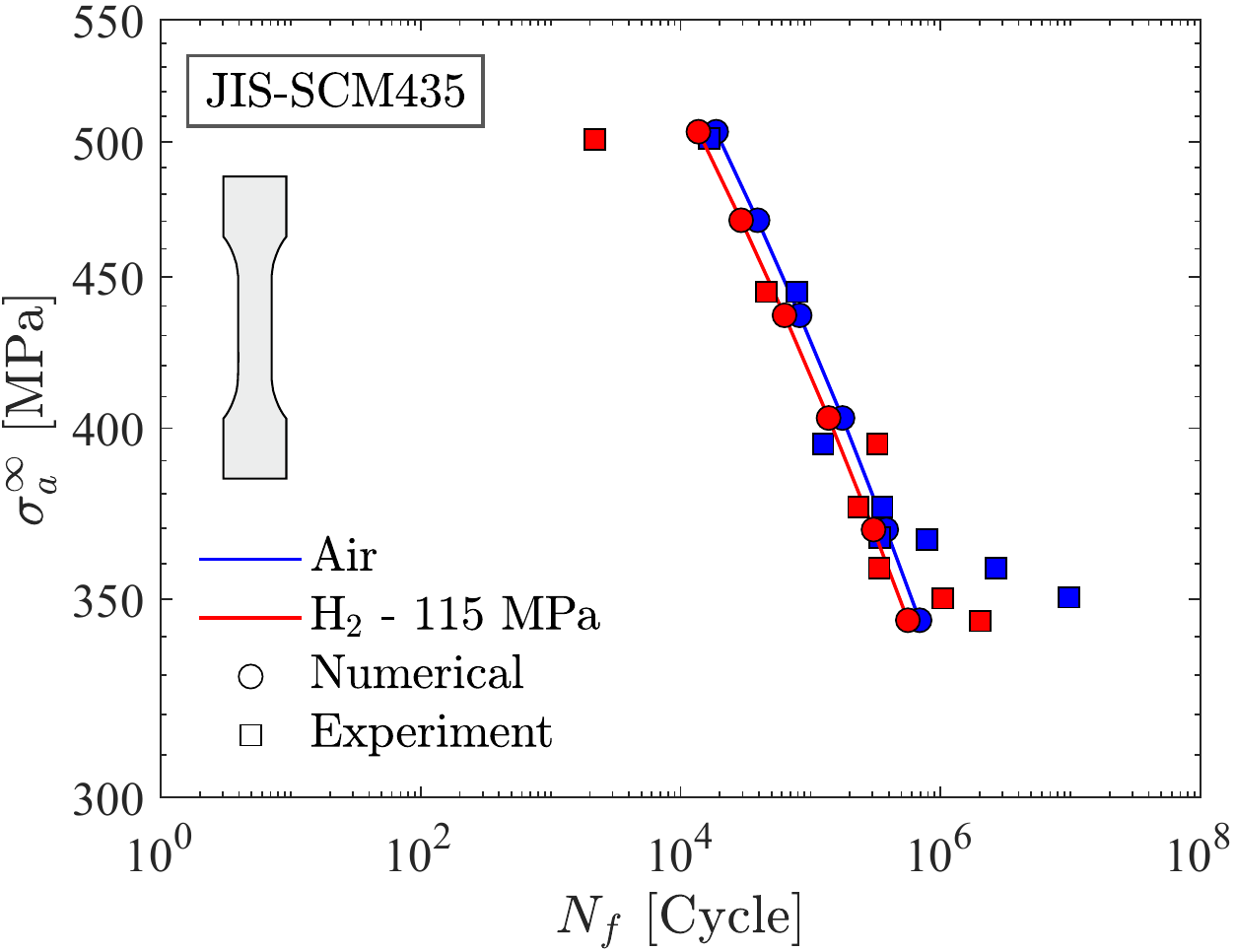}
    \caption{}
    \label{}
    \end{subfigure}
    \begin{subfigure}[t]{0.49\textwidth}
        \centering
    \includegraphics[width=1\linewidth,trim={0 0 0 0},clip]{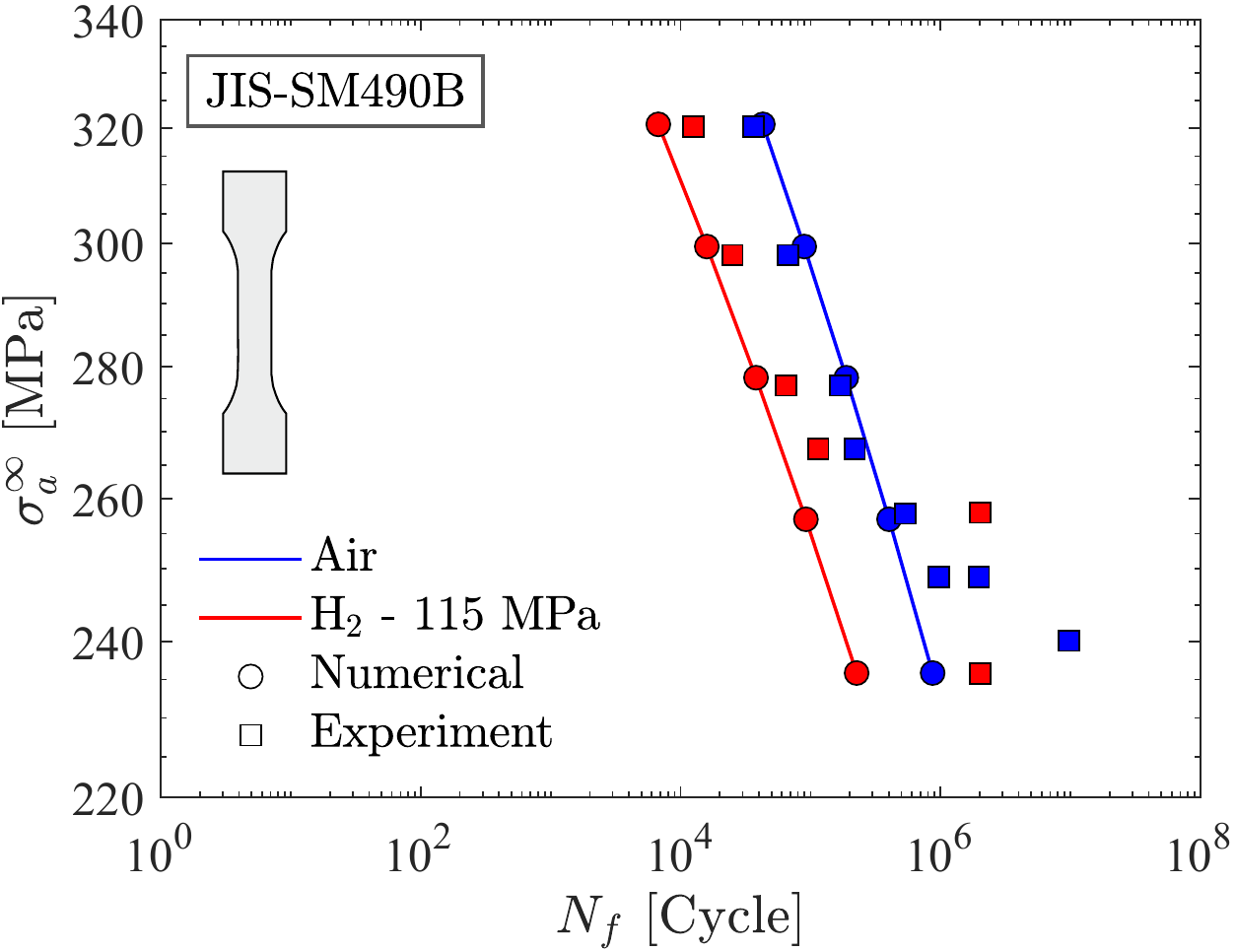}
    \caption{}
    \label{}    
    \end{subfigure}
    \caption{S-N curves from smooth samples: numerical (present) and experimental \cite{Matsunaga2015} in air and at hydrogen pressure of 115 MPa. Two materials are considered: (a) JIS-SCM435, a Cr-Mo steel, and (b) JIS-SM490B, a carbon steel.}
    \label{fig:SNcurves}
\end{figure}

The experimental and numerical results obtained are shown in Fig. \ref{fig:SNcurves}. Despite the scatter typically associated with these experiments, the \emph{Virtual} S-N curves predicted are in good agreement with the measured data. In both experiments and simulations, a higher susceptibility to hydrogen-assisted fatigue is observed in the case of JIS-SM490B, a steel with a higher solubility, where hydrogen reduces the number of cycles to failure by almost an order of magnitude. It is also worth noting that the agreement with experiments becomes less satisfactory at low stress amplitudes, particularly in the absence of hydrogen. This is likely to be improved if a fatigue endurance limit is incorporated into the modelling. Future work will be targeted towards this extension and the investigation of the role of hydrogen in the fatigue endurance of metals.

\section{Conclusions}
\label{Sec:Conclusions}

We have presented a multi-physics phase field-based model for hydrogen-assisted fatigue. Cracking is predicted with an energy based criterion grounded on the thermodynamics of crack growth, and the role of hydrogen is incorporated through a first-principles degradation of the fracture energy. Deformation, diffusion and fatigue crack growth are coupled, with the model capturing the solubility dependence on the hydrostatic stress and the evolving environment-diffusion interface. Several findings shall be emphasised:

\begin{itemize}
  \item The crack tip hydrogen distribution is very sensitive to the loading frequency $f$ and the material diffusivity $D$. Sufficiently high $f$ values lead to a hydrogen concentration that does not exhibit cyclic oscillations and increases in time up to a saturation value (even for a load ratio of $R=0$).
  \item The model adequately captures the sensitivity of fatigue crack growth rates to hydrogen content.
  \item The model naturally recovers the Paris law behaviour and thus can quantify the influence of hydrogen on the Paris law parameters.
  \item The sensitivity of crack growth rates to loading frequency is mapped, revealing two limit states, as observed experimentally, and predicting a smooth transition in-between.
  \item \emph{Virtual} S-N curves are obtained for various environments and both notched and smooth samples. Parameter-free predictions of the impact of hydrogen on the S-N curves reveal a promising agreement with experiments.
\end{itemize}

The theoretical and numerical framework presented provides a platform for addressing the long-standing challenge of predicting hydrogen-assisted fatigue failures. 

\section{Acknowledgements}
\label{Acknowledge of funding}
 A. Golahmar acknowledges financial support from Vattenfall Vindkraft A/S and Innovation Fund Denmark (grant 0153-00018B).
 E. Mart\'{\i}nez-Pa\~neda acknowledges financial support from the EPSRC (grant EP/V009680/1) and from the Royal Commission for the 1851 Exhibition (RF496/2018).



\bibliographystyle{elsarticle-num}
\bibliography{library}

\begin{thebibliography}{10}
\expandafter\ifx\csname url\endcsname\relax
  \def\url#1{\texttt{#1}}\fi
\expandafter\ifx\csname urlprefix\endcsname\relax\def\urlprefix{URL }\fi
\expandafter\ifx\csname href\endcsname\relax
  \def\href#1#2{#2} \def\path#1{#1}\fi

\bibitem{Martin2013}
M.~L. Martin, P.~Sofronis, I.~M. Robertson, T.~Awane, Y.~Murakami, {A
  microstructural based understanding of hydrogen-enhanced fatigue of stainless
  steels}, International Journal of Fatigue 57 (2013) 28--36.

\bibitem{Colombo2015}
C.~Colombo, G.~Fumagalli, F.~Bolzoni, G.~Gobbi, L.~Vergani, {Fatigue behavior
  of hydrogen pre-charged low alloy Cr-Mo steel}, International Journal of
  Fatigue 83 (2015) 2--9.

\bibitem{Yamabe2017}
J.~Yamabe, M.~Yoshikawa, H.~Matsunaga, S.~Matsuoka, {Hydrogen trapping and
  fatigue crack growth property of low-carbon steel in hydrogen-gas
  environment}, International Journal of Fatigue 102 (2017) 202--213.

\bibitem{Peral2019}
L.~B. Peral, A.~Zafra, S.~Blas{\'{o}}n, C.~Rodr{\'{i}}guez, J.~Belzunce,
  {Effect of hydrogen on the fatigue crack growth rate of quenched and tempered
  CrMo and CrMoV steels}, International Journal of Fatigue 120 (2019) 201--214.

\bibitem{Shinko2019}
T.~Shinko, G.~H{\'{e}}naff, D.~Halm, G.~Benoit, G.~Bilotta, M.~Arzaghi,
  {Hydrogen-affected fatigue crack propagation at various loading frequencies
  and gaseous hydrogen pressures in commercially pure iron}, International
  Journal of Fatigue 121 (2019) 197--207.

\bibitem{Ogawa2020}
Y.~Ogawa, K.~Umakoshi, M.~Nakamura, O.~Takakuwa, H.~Matsunaga,
  {Hydrogen-assisted, intergranular, fatigue crack-growth in ferritic iron:
  Influences of hydrogen-gas pressure and temperature variation}, International
  Journal of Fatigue 140 (2020) 105806.

\bibitem{Gangloff2003}
R.~P. Gangloff, {Hydrogen-assisted Cracking}, in: I.~Milne, R.~Ritchie,
  B.~Karihaloo (Eds.), Comprehensive Structural Integrity Vol. 6, Elsevier
  Science, New York, NY, 2003, pp. 31--101.

\bibitem{Murakami2006}
Y.~Murakami, H.~Matsunaga, {The effect of hydrogen on fatigue properties of
  steels used for fuel cell system}, International Journal of Fatigue 28~(11)
  (2006) 1509--1520.

\bibitem{Gangloff2012}
R.~P. Gangloff, B.~P. Somerday, {Gaseous Hydrogen Embrittlement of Materials in
  Energy Technologies}, Woodhead Publishing Limited, Cambridge, 2012.

\bibitem{AM2016}
E.~Mart{\'{i}}nez-Pa{\~{n}}eda, C.~F. Niordson, R.~P. Gangloff, {Strain
  gradient plasticity-based modeling of hydrogen environment assisted
  cracking}, Acta Materialia 117 (2016) 321--332.

\bibitem{Burns2016a}
J.~T. Burns, Z.~D. Harris, J.~D. Dolph, R.~P. Gangloff, {Measurement and
  Modeling of Hydrogen Environment-Assisted Cracking in a Ni-Cu-Al-Ti
  Superalloy}, Metallurgical and Materials Transactions A: Physical Metallurgy
  and Materials Science 47~(3) (2016) 990--997.

\bibitem{Novak2010}
P.~Novak, R.~Yuan, B.~P. Somerday, P.~Sofronis, R.~O. Ritchie, {A statistical,
  physical-based, micro-mechanical model of hydrogen-induced intergranular
  fracture in steel}, Journal of the Mechanics and Physics of Solids 58~(2)
  (2010) 206--226.

\bibitem{Ayas2014}
C.~Ayas, V.~S. Deshpande, N.~A. Fleck, {A fracture criterion for the notch
  strength of high strength steels in the presence of hydrogen}, Journal of the
  Mechanics and Physics of Solids 63~(1) (2014) 80--93.

\bibitem{Serebrinsky2004}
S.~Serebrinsky, E.~A. Carter, M.~Ortiz, {A quantum-mechanically informed
  continuum model of hydrogen embrittlement}, Journal of the Mechanics and
  Physics of Solids 52~(10) (2004) 2403--2430.

\bibitem{Yu2017}
H.~Yu, J.~S. Olsen, V.~Olden, A.~Alvaro, J.~He, Z.~Zhang, {Continuum level
  simulation of the grain size and misorientation effects on hydrogen
  embrittlement in nickel}, Engineering Failure Analysis 81 (2017) 79--93.

\bibitem{Elmukashfi2020}
E.~Elmukashfi, E.~Tarleton, A.~C.~F. Cocks, {A modelling framework for coupled
  hydrogen diffusion and mechanical behaviour of engineering components},
  Computational Mechanics 66 (2020) 189--220.

\bibitem{Anand2019}
L.~Anand, Y.~Mao, B.~Talamini, {On modeling fracture of ferritic steels due to
  hydrogen embrittlement}, Journal of the Mechanics and Physics of Solids 122
  (2019) 280--314.

\bibitem{CMAME2018}
E.~Mart{\'{i}}nez-Pa{\~{n}}eda, A.~Golahmar, C.~F. Niordson, {A phase field
  formulation for hydrogen assisted cracking}, Computer Methods in Applied
  Mechanics and Engineering 342 (2018) 742--761.

\bibitem{Duda2018}
F.~P. Duda, A.~Ciarbonetti, S.~Toro, A.~E. Huespe, {A phase-field model for
  solute-assisted brittle fracture in elastic-plastic solids}, International
  Journal of Plasticity 102 (2018) 16--40.

\bibitem{Wu2020b}
J.-Y. Wu, T.~K. Mandal, V.~P. Nguyen, {A phase-field regularized cohesive zone
  model for hydrogen assisted cracking}, Computer Methods in Applied Mechanics
  and Engineering 358 (2020) 112614.

\bibitem{JMPS2020}
P.~K. Kristensen, C.~F. Niordson, E.~Mart{\'{i}}nez-Pa{\~{n}}eda, {A phase
  field model for elastic-gradient-plastic solids undergoing hydrogen
  embrittlement}, Journal of the Mechanics and Physics of Solids 143 (2020)
  104093.

\bibitem{TAFM2020c}
P.~K. Kristensen, C.~F. Niordson, E.~Mart{\'{i}}nez-Pa{\~{n}}eda, {Applications
  of phase field fracture in modelling hydrogen assisted failures}, Theoretical
  and Applied Fracture Mechanics 110 (2020) 102837.

\bibitem{CS2020}
E.~Mart{\'{i}}nez-Pa{\~{n}}eda, Z.~D. Harris, S.~Fuentes-Alonso, J.~R. Scully,
  J.~T. Burns, {On the suitability of slow strain rate tensile testing for
  assessing hydrogen embrittlement susceptibility}, Corrosion Science 163
  (2020) 108291.

\bibitem{Castelluccio2018}
G.~M. Castelluccio, C.~B. Geller, D.~L. McDowell, {A rationale for modeling
  hydrogen effects on plastic deformation across scales in FCC metals},
  International Journal of Plasticity 111 (2018) 72--84.

\bibitem{Hosseini2021}
Z.~S. Hosseini, M.~Dadfarnia, A.~Nagao, M.~Kubota, B.~P. Somerday, R.~O.
  Ritchie, P.~Sofronis, {Modeling the Hydrogen Effect on the Constitutive
  Response of a Low Carbon Steel in Cyclic Loading}, Journal of Applied
  Mechanics, Transactions ASME 88~(3) (2021) 1--14.

\bibitem{Esaklul1983}
K.~A. Esaklul, A.~G. Wright, W.~W. Gerberich, {The effect of hydrogen induced
  surface asperities on fatigue crack closure in ultrahigh strength steel},
  Scripta Metallurgica 17~(9) (1983) 1073--1078.

\bibitem{Esaklul1983b}
K.~A. Esaklul, W.~W. Gerberich, {On the influence of internal hydrogen on
  fatigue thresholds of HSLA steel}, Scripta Metallurgica 17 (1983) 1079--1082.

\bibitem{AM2020}
R.~Fern{\'{a}}ndez-Sousa, C.~Beteg{\'{o}}n, E.~Mart{\'{i}}nez-Pa{\~{n}}eda,
  {Analysis of the influence of microstructural traps on hydrogen assisted
  fatigue}, Acta Materialia 199 (2020) 253--263.

\bibitem{Shinko2021}
T.~Shinko, D.~Halm, G.~Benoit, G.~H{\'{e}}naff, {Controlling factors and
  mechanisms of fatigue crack growth influenced by high pressure of gaseous
  hydrogen in a commercially pure iron}, Theoretical and Applied Fracture
  Mechanics 112 (2021) 102885.

\bibitem{Moriconi2014}
C.~Moriconi, G.~H{\'{e}}naff, D.~Halm, {Cohesive zone modeling of fatigue crack
  propagation assisted by gaseous hydrogen in metals}, International Journal of
  Fatigue 68 (2014) 56--66.

\bibitem{EFM2017}
S.~del Busto, C.~Beteg{\'{o}}n, E.~Mart{\'{i}}nez-Pa{\~{n}}eda, {A cohesive
  zone framework for environmentally assisted fatigue}, Engineering Fracture
  Mechanics 185 (2017) 210--226.

\bibitem{Zhao2016}
Y.~Zhao, B.~X. Xu, P.~Stein, D.~Gross, {Phase-field study of electrochemical
  reactions at exterior and interior interfaces in Li-ion battery electrode
  particles}, Computer Methods in Applied Mechanics and Engineering 312 (2016)
  428--446.

\bibitem{Mesgarnejad2019}
A.~Mesgarnejad, A.~Karma, {Phase field modeling of chemomechanical fracture of
  intercalation electrodes: Role of charging rate and dimensionality}, Journal
  of the Mechanics and Physics of Solids 132 (2019).

\bibitem{Quintanas-Corominas2020a}
A.~Quintanas-Corominas, A.~Turon, J.~Reinoso, E.~Casoni, M.~Paggi, J.~A.
  Mayugo, {A phase field approach enhanced with a cohesive zone model for
  modeling delamination induced by matrix cracking}, Computer Methods in
  Applied Mechanics and Engineering 358 (2020) 112618.

\bibitem{CST2021}
W.~Tan, E.~Mart{\'{i}}nez-Pa{\~{n}}eda, {Phase field predictions of microscopic
  fracture and R-curve behaviour of fibre-reinforced composites}, Composites
  Science and Technology 202 (2021) 108539.

\bibitem{Carollo2018}
V.~Carollo, J.~Reinoso, M.~Paggi, {Modeling complex crack paths in ceramic
  laminates: A novel variational framework combining the phase field method of
  fracture and the cohesive zone model}, Journal of the European Ceramic
  Society 38~(8) (2018) 2994--3003.

\bibitem{Li2021}
W.~Li, K.~Shirvan, {Multiphysics phase-field modeling of quasi-static cracking
  in urania ceramic nuclear fuel}, Ceramics International 47 (2021) 793--810.

\bibitem{CMAME2021}
M.~Simoes, E.~Mart{\'{i}}nez-Pa{\~{n}}eda, {Phase field modelling of fracture
  and fatigue in Shape Memory Alloys}, Computer Methods in Applied Mechanics
  and Engineering 373 (2021) 113504.

\bibitem{CPB2019}
Hirshikesh, S.~Natarajan, R.~K. Annabattula, E.~Mart{\'{i}}nez-Pa{\~{n}}eda,
  {Phase field modelling of crack propagation in functionally graded
  materials}, Composites Part B: Engineering 169 (2019) 239--248.

\bibitem{Kumar2021}
P.~K. A.~V. Kumar, A.~Dean, J.~Reinoso, P.~Lenarda, M.~Paggi, {Phase field
  modeling of fracture in Functionally Graded Materials: G -convergence and
  mechanical insight on the effect of grading}, Thin-Walled Structures 159
  (2021) 107234.

\bibitem{McAuliffe2015}
C.~McAuliffe, H.~Waisman, {A unified model for metal failure capturing shear
  banding and fracture}, International Journal of Plasticity 65 (2015)
  131--151.

\bibitem{Borden2016}
M.~J. Borden, T.~J.~R. Hughes, C.~M. Landis, A.~Anvari, I.~J. Lee, {A
  phase-field formulation for fracture in ductile materials: Finite deformation
  balance law derivation, plastic degradation, and stress triaxiality effects},
  Computer Methods in Applied Mechanics and Engineering 312 (2016) 130--166.

\bibitem{IJP2021}
M.~Isfandbod, E.~Mart{\'{i}}nez-Pa{\~{n}}eda, {A mechanism-based multi-trap
  phase field model for hydrogen assisted fracture}, International Journal of
  Plasticity 144 (2021) 103044.

\bibitem{Francfort1998}
G.~A. Francfort, J.-J. Marigo, {Revisiting brittle fracture as an energy
  minimization problem}, Journal of the Mechanics and Physics of Solids 46~(8)
  (1998) 1319--1342.

\bibitem{Bourdin2000}
B.~Bourdin, G.~A. Francfort, J.-J. Marigo, {Numerical experiments in revisited
  brittle fracture}, Journal of the Mechanics and Physics of Solids 48~(4)
  (2000) 797--826.

\bibitem{Provatas2011}
N.~Provatas, K.~Elder, {Phase-Field Methods in Materials Science and
  Engineering}, John Wiley {\&} Sons, Weinheim, Germany, 2011.

\bibitem{JMPS2021}
C.~Cui, R.~Ma, E.~Mart{\'{i}}nez-Pa{\~{n}}eda, {A phase field formulation for
  dissolution-driven stress corrosion cracking}, Journal of the Mechanics and
  Physics of Solids 147 (2021) 104254.

\bibitem{Chambolle2004}
A.~Chambolle, {An approximation result for special functions with bounded
  deformation}, Journal des Mathematiques Pures et Appliquees 83~(7) (2004)
  929--954.

\bibitem{Tanne2018}
E.~Tann{\'{e}}, T.~Li, B.~Bourdin, J.-J. Marigo, C.~Maurini, {Crack nucleation
  in variational phase-field models of brittle fracture}, Journal of the
  Mechanics and Physics of Solids 110 (2018) 80--99.

\bibitem{PTRSA2021}
P.~K. Kristensen, C.~F. Niordson, E.~Mart{\'{i}}nez-Pa{\~{n}}eda, {An
  assessment of phase field fracture: crack initiation and growth},
  Philosophical Transactions of the Royal Society A: Mathematical, Physical and
  Engineering Sciences 379 (2021) 20210021.

\bibitem{Jiang2004a}
D.~E. Jiang, E.~A. Carter, {First principles assessment of ideal fracture
  energies of materials with mobile impurities: Implications for hydrogen
  embrittlement of metals}, Acta Materialia 52~(16) (2004) 4801--4807.

\bibitem{Carrara2020}
P.~Carrara, M.~Ambati, R.~Alessi, L.~{De Lorenzis}, {A framework to model the
  fatigue behavior of brittle materials based on a variational phase-field
  approach}, Computer Methods in Applied Mechanics and Engineering 361 (2020)
  112731.

\bibitem{Miehe2010a}
C.~Miehe, M.~Hofacker, F.~Welschinger, {A phase field model for
  rate-independent crack propagation: Robust algorithmic implementation based
  on operator splits}, Computer Methods in Applied Mechanics and Engineering
  199~(45-48) (2010) 2765--2778.

\bibitem{Ambati2015}
M.~Ambati, T.~Gerasimov, L.~{De Lorenzis}, {A review on phase-field models of
  brittle fracture and a new fast hybrid formulation}, Computational Mechanics
  55 (2015) 383--405.

\bibitem{Amor2009}
H.~Amor, J.~J. Marigo, C.~Maurini, {Regularized formulation of the variational
  brittle fracture with unilateral contact: Numerical experiments}, Journal of
  the Mechanics and Physics of Solids 57~(8) (2009) 1209--1229.

\bibitem{Seles2019a}
K.~Sele{\v{s}}, T.~Lesi{\v{c}}ar, Z.~Tonkovi{\'{c}}, J.~Sori{\'{c}}, {A
  residual control staggered solution scheme for the phase-field modeling of
  brittle fracture}, Engineering Fracture Mechanics 205 (2019) 370--386.

\bibitem{Renard2020}
Y.~Renard, K.~Poulios, {GetFEM: Automated FE modeling of multiphysics problems
  based on a generic weak form language}, ACM Transactions on Mathematical
  Software (TOMS) 47~(1) (2020) 1--31.

\bibitem{TAFM2020}
P.~K. Kristensen, E.~Mart{\'{i}}nez-Pa{\~{n}}eda, {Phase field fracture
  modelling using quasi-Newton methods and a new adaptive step scheme},
  Theoretical and Applied Fracture Mechanics 107 (2020) 102446.

\bibitem{Turnbull1996}
A.~Turnbull, D.~H. Ferriss, H.~Anzai, {Modelling of the hydrogen distribution
  at a crack tip}, Materials Science and Engineering A 206~(1) (1996) 1--13.

\bibitem{CS2020b}
E.~Mart{\'{i}}nez-Pa{\~{n}}eda, A.~D{\'{i}}az, L.~Wright, A.~Turnbull,
  {Generalised boundary conditions for hydrogen transport at crack tips},
  Corrosion Science 173 (2020) 108698.

\bibitem{DiLeo2013}
C.~V. {Di Leo}, L.~Anand, {Hydrogen in metals: A coupled theory for species
  diffusion and large elastic-plastic deformations}, International Journal of
  Plasticity 43 (2013) 42--69.

\bibitem{IJHE2016}
E.~Mart{\'{i}}nez-Pa{\~{n}}eda, S.~del Busto, C.~F. Niordson, C.~Beteg{\'{o}}n,
  {Strain gradient plasticity modeling of hydrogen diffusion to the crack tip},
  International Journal of Hydrogen Energy 41~(24) (2016) 10265--10274.

\bibitem{Diaz2016b}
A.~D{\'{i}}az, J.~M. Alegre, I.~I. Cuesta, {Coupled hydrogen diffusion
  simulation using a heat transfer analogy}, International Journal of
  Mechanical Sciences 115-116 (2016) 360--369.

\bibitem{Gangloff1990}
R.~P. Gangloff, {Corrosion fatigue crack propagation in metals}, Tech. rep.,
  NASA 19900015089 (1990).

\bibitem{Williams1957}
M.~L. Williams, {On the stress distribution at the base of a stationary crack},
  Journal of Applied Mechanics 24 (1957) 109--114.

\bibitem{Anderson2005}
T.~L. Anderson, {Fracture Mechanics. Fundamentals and Applications}, 3rd
  Edition, CRC Press, Taylor {\&} Francis, Boca Raton, 2005.

\bibitem{Matsunaga2015}
H.~Matsunaga, M.~Yoshikawa, R.~Kondo, J.~Yamabe, S.~Matsuoka, {Slow strain rate
  tensile and fatigue properties of Cr-Mo and carbon steels in a 115 MPa
  hydrogen gas atmosphere}, International Journal of Hydrogen Energy 40~(16)
  (2015) 5739--5748.

\bibitem{Matsumoto2017}
T.~Matsumoto, M.~Kubota, S.~Matsuoka, P.~Ginet, J.~Furtado, F.~Barbier,
  {Threshold stress intensity factor for hydrogen-assisted cracking of CR-MO
  steel used as stationary storage buffer of a hydrogen refueling station},
  International Journal of Hydrogen Energy 42~(11) (2017) 7422--7428.

\bibitem{Ogawa2017}
Y.~Ogawa, H.~Matsunaga, J.~Yamabe, M.~Yoshikawa, S.~Matsuoka, {Unified
  evaluation of hydrogen-induced crack growth in fatigue tests and fracture
  toughness tests of a carbon steel}, International Journal of Fatigue 103
  (2017) 223--233.

\bibitem{SanMarchi2012}
C.~{San Marchi}, B.~P. Somerday, {Technical Reference for Hydrogen
  Compatibility of Materials}, Tech. rep., SANDIA National Labs (2012).

\end{thebibliography}
\end{document}